\documentclass[12pt]{article}

\usepackage{graphicx}
\usepackage[dvips]{color}

\usepackage{amsmath,amssymb,epsf,epic}

\newcommand{\un}{{\mathbb I}}
\newcommand{\ra}{\rightarrow}

\newcommand{\bra}{\langle} 

\newcommand{\ket}{\rangle}

\newcommand{\E}{{\mathbb E}}
\newcommand{\D}{{\mathbb D}}

\newcommand{\be}{\begin{equation}}
\newcommand{\ee}{\end{equation}}
\newcommand{\bea}{\begin{eqnarray}}
\newcommand{\eea}{\end{eqnarray}}

\newcommand{\eps}{\epsilon}

\newcommand{\ffi}{\varphi}

\newcommand{\ep}{\hfill  {\vrule height 10pt width 8pt depth 0pt}}

\newcommand{\grintl}{[\kern-.18em [}
\newcommand{\grintr}{]\kern-.18em ]}

\newcounter{resultcounter}[section]

\newtheorem{thm}[resultcounter]{Theorem}
\newtheorem{lem}[resultcounter]{Lemma}
\newtheorem{prop}[resultcounter]{Proposition}
\newtheorem{cor}[resultcounter]{Corollary}

\newtheorem{rem}[resultcounter]{Remark}
\newtheorem{rems}[resultcounter]{Remarks}

  \def\cC{{\cal C}}
  
\def\cG{{\cal G}} \def\cH{{\cal H}} 
 \def\cK{{\cal K}} 
\def\cM{{\cal M}} \def\cN{{\cal N}} 
\def\cP{{\cal P}}  \def\cR{{\cal R}}
\def\cS{{\cal S}} \def\cT{{\cal T}}

\newcommand{\R}{{\mathbb R}}
\newcommand{\N}{{\mathbb N}}

\newcommand{\C}{{\mathbb C}}
\newcommand{\Z}{{\mathbb Z}}
\renewcommand{\E}{{\mathbb E}}
\renewcommand{\P}{{\mathbb P}}
\newcommand{\I}{{\mathbb I}}
\newcommand{\T}{{\mathbb T}}

\newcommand{\U}{{\mathbb U}}

\begin{document}
\title{{Spectral Transition} for Random Quantum Walks on Trees}
 \author{ Eman Hamza\footnote{Department of Physics, Faculty of Science, Cairo University,Cairo 12613, Egypt} \footnote{ Partially supported by a Fulbright research grant}\and Alain Joye\footnote{ UJF-Grenoble 1, CNRS Institut Fourier UMR 5582, Grenoble, 38402, France} \footnote{Partially supported by the Agence Nationale de la Recherche, grant ANR-09-BLAN-0098-01}}

\date{ }

\maketitle
\vspace{-1cm}

\thispagestyle{empty}
\setcounter{page}{1}
\setcounter{section}{1}

\setcounter{section}{0}

\abstract{ We define and analyze random quantum walks on homogeneous trees of degree $q\geq 3$. Such walks describe the discrete time evolution of a quantum particle with internal degree of freedom in $\C^q$ hopping on the neighboring sites of the tree in presence of static disorder. The one time step random unitary evolution operator of the particle depends on a unitary matrix $C\in U(q)$ which monitors the strength of the disorder. We prove for any $q$ that there exist open sets of matrices in $U(q)$ for which the random evolution has either pure point spectrum almost surely or purely absolutely continuous spectrum, thereby showing the existence of a spectral transition driven by $C\in U(q)$. For $q\in\{3,4\}$, we establish properties of the spectral diagram which provide a description of the spectral transition.
}

\thispagestyle{empty}
\setcounter{page}{1}
\setcounter{section}{1}

\setcounter{section}{0}

\section{Introduction}

 {
 
Quantum walks 
have become a popular research topic in the recent years due to the role they play in several different fields, see for example the reviews \cite{Ke, Ko, V-A} and references therein. They are typically defined as discrete time quantum dynamical systems characterized by a unitary operator on the Hilbert space of a particle with internal degree of freedom on a graph with the proviso that only neighboring sites are coupled by the unitary operator. Quantum walks are used to approximate the dynamics of certain quantum systems in appropriate regimes: 
electrons in a two dimensional random potential and a large 
magnetic field, atoms trapped in some time dependent optical lattices, ions caught in suitably tuned Paul magnetic traps or polarized photons propagating  in networks of imperfect waveguides display dynamics that can be described by quantum walks on graphs, \cite{CC, Ketal, Zetal, sciarrino}.
 In the quantum computing community, the algorithmic simplicity of quantum walks provides them with a distinguished role. They are used as tools assessing the probabilistic efficiency of quantum search  algorithms to be implemented on quantum computers and 
 they also provide building blocks in the elaboration of such algorithms, see e.g. \cite{S, MNRS}. 
Depending on the framework, several variants of quantum walks are considered: completely positive maps can be used to extend the unitary setup \cite{AAKV, Gu, APSS}, the stationarity assumption can be relaxed allowing one to deal with  genuinely time dependent walks \cite{AVWW, J2, HJ} or the deterministic framework can be enlarged to accommodate random evolution operators from a  set of unitary operators \cite{CC, KLMW, J4}. The latter are called {\it random quantum walks} and they describe the motion of a quantum walker in a static random environment. 

In this paper we define and analyze  
random quantum walks describing the dynamics of a quantum particle with internal degree of freedom hopping on homogeneous trees of degree $q$, $q\geq 3$, in a static random environment. The internal degree of freedom, or coin state, lives in $\C^q$. The deterministic part of the walk is given by a {\em coined quantum walk}: the one time step unitary evolution $U(C)$ is obtained by the action of a unitary matrix $C\in U(q)$ on the coin state of the particle, followed by the action of a coin state conditioned shift $S$ which moves the particle to its nearest neighbors on the tree. 
Static disorder is introduced 
via {\em  i.i.d. random phases} used to decorate the coin matrix $C$ in such a way that the unitary coin state update becomes random and {\em site-dependent on the tree}. The coin matrix $C$ is regarded as a parameter of the resulting random unitary operator $U_\omega(C)$, see the precise definition in the next section. 
 Our definition of quantum walks on 
 trees differs from those available in the literature, see {\it e.g.} \cite{CHKS, Detal}, in that the repeated action of the coin state conditioned shift $S$ alone actually makes the quantum walker propagate on the tree.

We analyze the spectrum of the random evolution $U_\omega(C)$ as a function of $C$ which, in analogy with the self-adjoint Anderson model, we consider as a $U(q)$ valued parameter monitoring the {\em strength of the disorder}. The mechanism of Anderson localization is expected to produce regimes of complete suppression of transport and to pure point spectrum.
Spectral and dynamical localization have been proven to hold for random quantum walks analogous to $U_\omega(C)$ defined on $\Z^d$, in a large disorder regime and at the band edges for arbitrary disorder strength for $d\geq 2$, and  for any disorder when $d=1$. See \cite{J1, HJS1, HJS2, ABJ2, JM, ASWe, J3}. 
For random quantum walks on trees, spectral {\em delocalization} at weak disorder and spectral {\em localization} at large disorder are expected, by analogy with the self-adjoint case. For the Anderson model on the Bethe lattice, this spectral transition is an 
established fact,  the detailed analysis of which is the object of ongoing investigations, see e.g. \cite{Kl, AW1, AW2}. 
We show that for random quantum walks on trees of degree $q$, the spectral nature of $U_\omega(C)$ depend crucially on the parameter $C\in U(q)$, and that a similar picture holds. First, we prove that for any $q$, there exist distinct open sets of $U(q)$ 
which determine regimes of coin matrices for which either $\sigma(U_\omega(C))$ is pure point almost surely, see Theorem \ref{fme} and Corollary \ref{cor1}, or  $\sigma(U_\omega(C))$ is purely absolutely continuous, see Propositions \ref{weakdelocodd} and \ref{weakdeloceven}. This establishes that a spectral transition driven by $C$ takes place, since $U(q)$ is compact and connected. Second, we discuss the salient features of the spectral diagrams for $q\in \{3, 4\}$, as illustrated in Figures  \ref{phase} and \ref{phase4}. In particular, we exhibit continuous families of coin matrices which interpolate between the localizing and delocalizing regimes along which we provide a complete description of the localization-delocalization transition.

The next section provides the definitions of random coined quantum walks on trees 
followed by a description of the spectral criteria suited to the present framework. 
Localization is proven by means of the fractional moments method in Section \ref{secloc}, whereas delocalization is a consequence of  dynamical spectral criteria described in Section \ref{secdeloc}.  Finally, Section \ref{q=3q=4} is devoted to a detailed analysis of the spectral transition in the cases $q=3, 4$.

 }

{\bf Acknowledgments } E.H. wishes to thank the ANR Ham Mark and Universit\'e Grenoble 1 for support in the Spring of 2012, where this work was initiated.   E.H. is also very grateful for the hospitality at the University of California, Davis during a sabbatical leave.

\section{General Setup}\label{setup}

\begin{figure}[htbp]
   \begin{center}
\begin{minipage}{.4\textwidth}
\begin{center}
      \includegraphics[scale=.4]{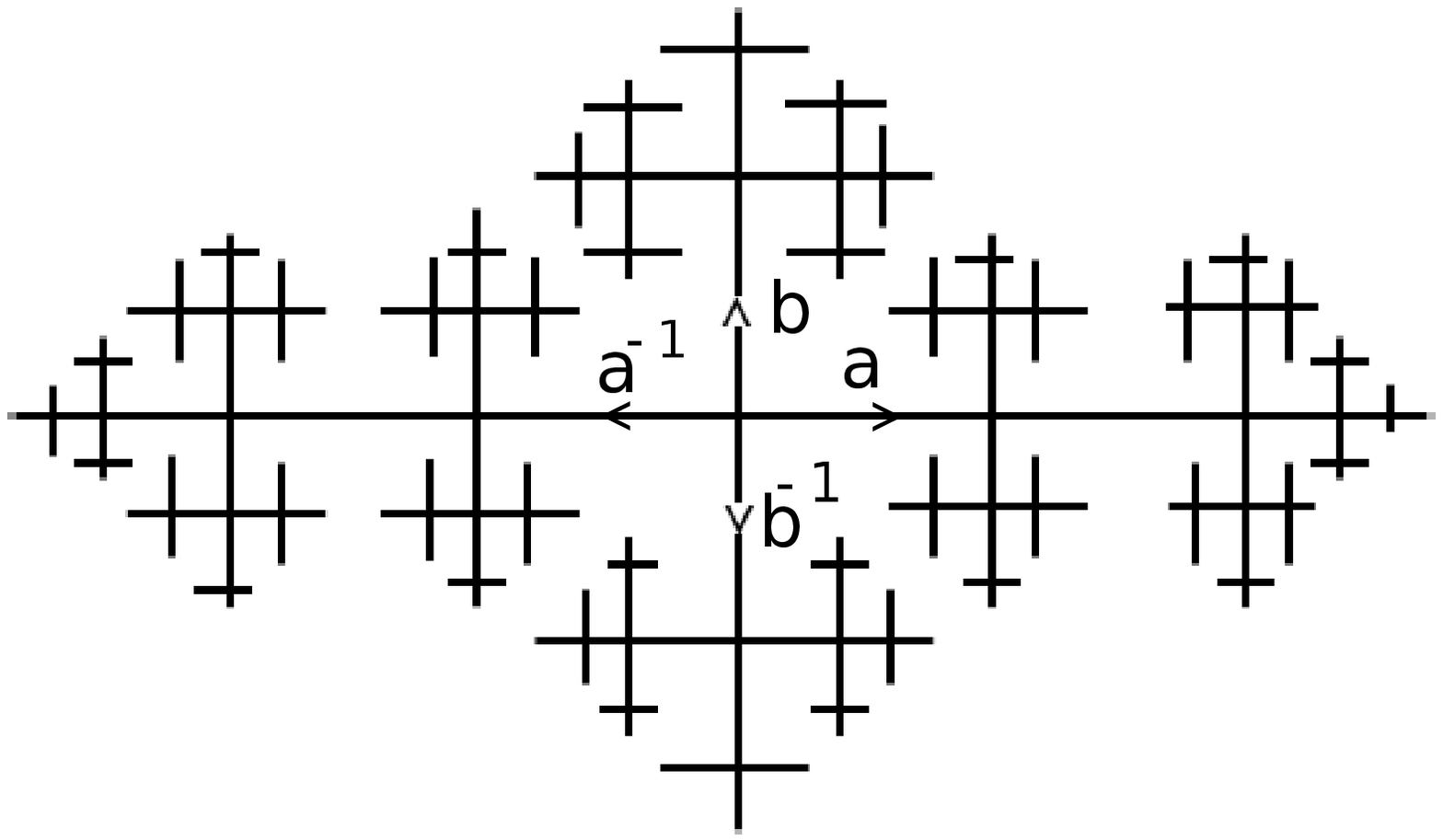}
   \end{center}
   \caption{\footnotesize construction of $\cT_4$}\label{tree4}
\end{minipage}
\begin{minipage}{.4\textwidth}
 \begin{center}
      \includegraphics[scale=.4]{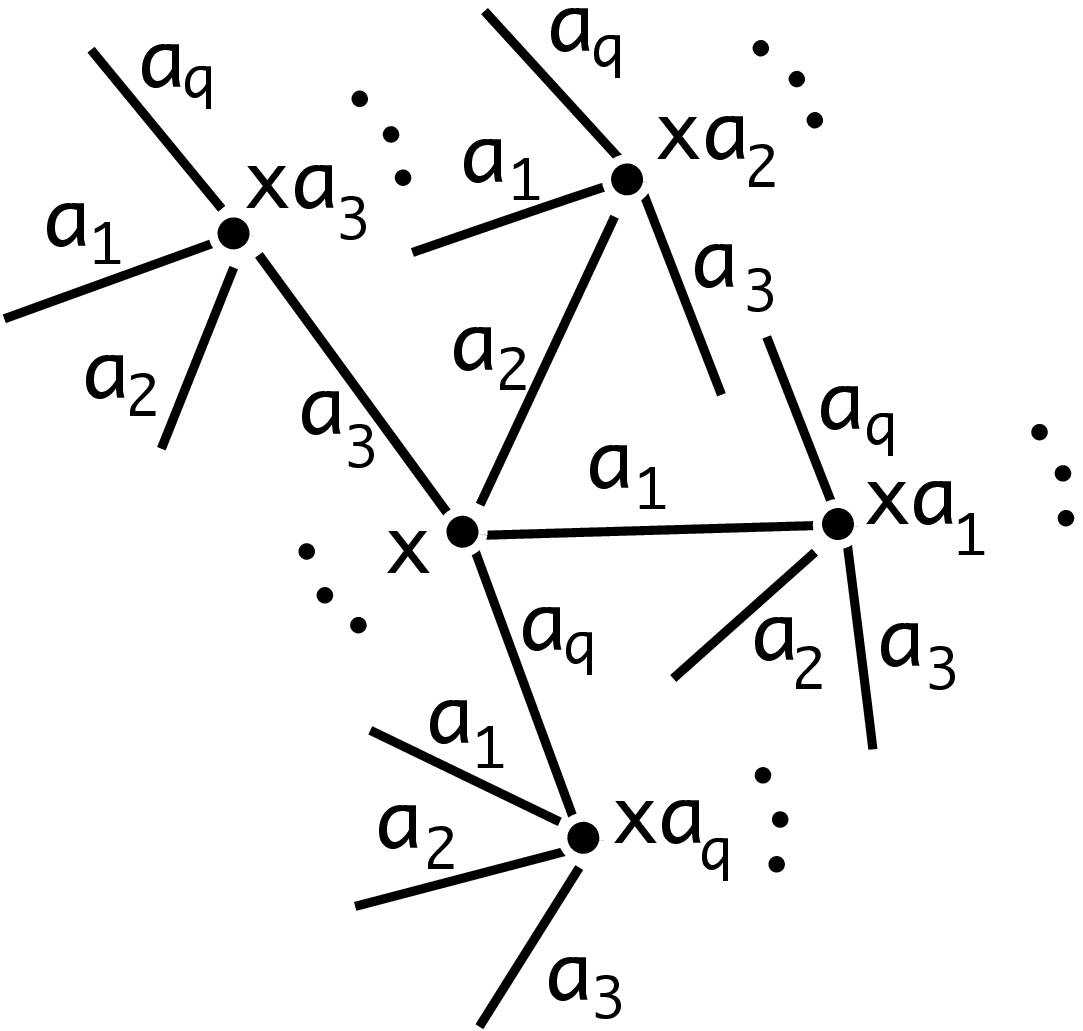}
   \end{center}
   \caption{\footnotesize Construction of $\cT_q$, with $q$ odd.}\label{tqodd}
\end{minipage}
\end{center}
\end{figure}

Let $\cT_q$ be a homogeneous tree of degree $q\geq 3$. If $q$ is even, we consider $\cT_q$ as the tree of the free group generated by 
\be
A_q=\{a_1, a_2, \dots, a_{q}\}\equiv\{a_1, a_2, \dots, a_{q/2}, a_1^{-1}, a_2^{-1}, \dots, a_{q/2}^{-1}\}
\ee 
with $a_ja_j^{-1}=a_j^{-1}a_j=e$, $e$ being the neutral element of the group; see Figure (\ref{tree4}) for $q=4$. If $q$ is odd, $\cT_q$ is considered as the tree generated by 
\be
A_q=\{a_1, a_2, \dots, a_{q}\} \ \mbox{such that $a_j^2=e$}.
\ee

We choose a vertex of $\cT_q$ to be the root of the tree, denoted by $e$. Each vertex $x=x_{1}x_{2}\dots x_{n}$, $n\in\N$ of $\cT_q$ is a reduced word made of finitely many letters from the alphabet $A_q$. An edge of $\cT_q$ consists in a pair of vertices $(x,y)$ such that $xy^{-1}\in A_q$. This defines nearest neighbors in $\cT_q$ and the number of nearest neighbors of any vertex is $q$. Any pair of vertices $x$ and $y$ can be joined by a unique set of edges, or path in $\cT_q$. The distance $|x|$ of a vertex $x=x_{1}x_{2}\dots x_{n}$ to the root  is $n$  and we denote by $d(x,y)$ the distance between two arbitrary vertices.

When $q$ is odd, the edges going away from a vertex $x$ are labeled as in Figure \ref{tqodd}:
Given the order $A_q=\{a_1, a_2, \dots, a_{q}\}$, the sequence of nearest neighbors of $x$,  $xa_j$, for $j=1,\dots, q$ are ordered around $x$ in the positive orientation. Then, the nearest neighbors of each $xa_j$ are arranged in the same order in such a way that the edge between $xa_j$ and $x$ corresponds to $a_j$.
Identifying $\cT_q$ with its set of vertices, the configuration Hilbert space of the walker is defined as
\be
l^2(\cT_q)=\Big\{\psi=\sum_{x\in \cT_q}\psi_x |x\ket \ \mbox{s.t.} \  \psi_x\in \C,  \ \sum_{x\in \cT_q}|\psi_x|^2<\infty\Big\},
\ee
where $|x\ket$ denotes the element of the canonical basis of $l^2(\cT_q)$ which sits at vertex $x$.
The coin Hilbert space of our quantum walker on $\cT_q$ is $\C^q$. It allows us to label the elements of the canonical basis of $\C^q$ by means of the letters of the alphabet $A_q$ as $\{|a_j\ket\in \C^q\}_{j=1,\dots,q}$.
Altogether, the total Hilbert space is
\be\label{canbas}
\cK_q = l^2(\cT_q)\otimes \C^q \ \mbox{ with canonical basis }\  \big\{x\otimes a\equiv |x\ket\otimes |a\ket ,\ \ x\in \cT_q, a\in A_q\big\}.
\ee
\begin{rem}
The dimension $q$ of the coin space is the smallest choice allowed by the condition that our quantum walk operator couples nearest neighbors on $\cT_q$ only.
\end{rem}
The dynamics we consider is defined as the composition of a unitary update of the coin variables in $\C^q$ followed by a coin state dependent shift on the tree. 
Let $C\in U(q)$, the set of $q\times q$ unitary matrices. The unitary update operator given by $\I\otimes C$ acts on the canonical basis of $\cK_q$ as
\be\label{reshuffle}
(\I\otimes C) x\otimes a=|x\ket\otimes |Ca\ket=\sum_{b\in A_q} C_{ba}\, x\otimes b,
\ee
where $\{C_{ba}\}_{(b,a)\in A_q^2}$ denote the matrix elements of $C$. 
The definition of the coin state dependent shift $S$ depends on the parity of $q$.

When $q$ is even, the unitary shift operator $S$ on $\cK_q$ is defined  by
\bea\label{dirsumshift}
S={\sum}_{a\in A_q} S_a\otimes |a\ket\bra a| 
={\sum_{a\in A_q\atop x\in \cT_q} |xa\ket\bra x |\otimes |a\ket\bra a| , }
\eea
where for all $a\in A_q$ the unitary operator $S_a$ acts on $l^2(\cT_q)$ as  
$
S_a|x\ket=|xa\ket,  \forall x\in \cT_q
$,
and $S_a^{-1}=S_a^*=S_{a^{-1}}$. Also
if $\cH^a_x\subset l^2(\cT_q)$ denotes the $S_a$-cyclic subspace generated by $|x\ket$,
\be
\cH^{a}_x=\overline{\mbox{span }}\big\{S_{a}^n |x\ket, \ n\in\Z\big\}=\overline{\mbox{span }}\big\{\cdots, |xa^{-1}a^{-1}\ket, |xa^{-1}\ket, |x\ket, |xa\ket, |xaa\ket, \cdots\big\}
\ee
(where the notation $\overline{\mbox{span }}$ means the closure of the span of the vectors considered)
then $\cH^a_x$ is isomorphic to $l^2(\Z)$ and $S_{a}|_{\cH^{a}_x}$ is unitarily equivalent to the shift on $l^2(\Z)$.
We define  the one step unitary evolution operator on $\cH=l^2(\cT_q)\otimes \C^q$ for $q$ even by 
\be\label{defeven}
U(C)=S(\I\otimes C)={\sum}_{a\in A_q} S_a\otimes |a\ket\bra a| C.
\ee

\begin{figure}[htbp]
   \begin{center}
      \includegraphics[scale=.28, angle=-90]{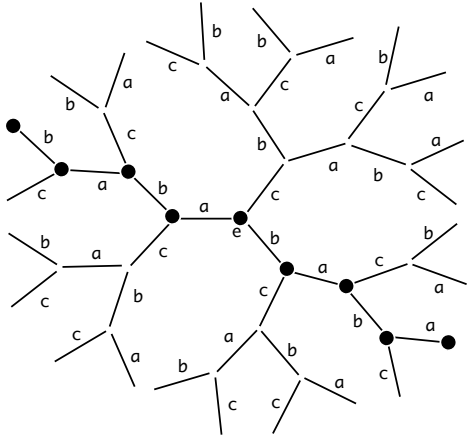}
   \end{center}
   \caption{\footnotesize The sites $\{S^{n}_{ab}e\}_{n\in\Z}$, $q=3$.}\label{figshift}
\end{figure}

When $q$ is odd, we construct a shift operator $S$ on $\cK_q=l^2(\cT_q)\otimes \C^q$ as a direct sum similar to (\ref{dirsumshift}) as follows.
Let $x_e$, respectively $x_o$, denote vertices of even, respectively odd length. Such vertices will be called odd sites, respectively even sites in the sequel. For $a\neq b\in A_q$, we define $S_{ab}$ on $l^2(\cT_q)$ by
\be
{S_{ab}=\sum_{x_e \in \cT_q} |x_ea\ket\bra x_e |+ \sum_{x_o \in \cT_q} |x_o b\ket\bra x_o |. }
\ee
Thus $S_{ab}^*=S_{ab}^{-1}=S_{ba}$ and 
$S_{ab}S_{cd}|x_e\ket =|x_ecb\ket,$ $S_{ab}S_{cd}|x_o\ket=|x_o da\ket,$
for all $a\neq b,c\neq d\in A_q$.
For each $x\in \cT_q$, let $\cH^{ab}_x$ the $S_{ab}$-cyclic subspace generated by $|x\ket$, 
\be
\cH^{ab}_x=\overline{\mbox{span }}\big\{S_{ab}^n |x\ket, \ n\in\Z\big\}\subset l^2(\cT_q).
\ee
See Figure \ref{figshift} for the sites of $\cH^{ab}_e$ on $\cT_3$,  (and Section \ref{q=3q=4} for the notation). One checks that
\begin{lem}
The subspace $\cH^{ab}_x$ is isomorphic to $l^2(\Z)$ and $S_{ab}|_{\cH^{ab}_x}$ is unitarily equivalent to the shift on $l^2(\Z)$.
\end{lem}
To define $S$, we make use of the $q$ shifts 
$
S_{a_1a_2}, S_{a_2a_3}, \cdots, S_{a_{q-1}a_q}, S_{a_qa_1}
$
only.
Let $A_q=\{a_1, a_2, \dots, a_{q}\}$, and let us denote by $\{|a_j\ket\}_{j=1,2,\dots,q}$ the elements of the canonical basis of $\C^q$. If need be, we write $a_{q+k}=a_k$, $k=1,2,\dots, q$. 
Then $S:\cK_q \ra \cK_q$ is given by
\bea
S&=&
{\sum_{1\leq j \leq q}} S_{a_{j+1}a_{j+2}}\otimes |a_j\ket \bra a_j|
\eea
and the one step unitary evolution operator on $\cK_q=l^2(T_q)\otimes \C^{q}$ for $q$ odd is defined by 
\be\label{defodd}
U(C)=S(\I\otimes C)={\sum_{1\leq j \leq q}} S_{a_{j+1}a_{j+2}}\otimes |a_j\ket \bra a_j|C.
\ee

A natural generalization consists in considering families of coin matrices $\cC=\{C(x)\in U(q)\}_{x\in\cT_q}$, indexed by the vertices $x\in\cT_q$ .
A quantum walk with site dependent coin matrices is defined through the formulas, 
for $q$ even, respectively $q$ odd,
\bea\label{explicitsitedep}
U(\cC)&=&\sum_{a\in A_q\atop x\in \cT_q} |xa\ket\bra x |\otimes |a\ket\bra a|  C(x), \ \ \mbox{respectively}\\ \nonumber
U(\cC)&=&\sum_{1\leq j \leq q} \left(\sum_{x_e \in \cT_q} |x_ea_{j+1}\ket\bra x_e |\otimes |a_j\ket \bra a_j| C(x_e) +\sum_{x_o \in \cT_q} |x_oa_{j+2}\ket\bra x_o |\otimes |a_j\ket \bra a_j| C(x_o) \right).
\eea
We deal with
 families of site dependent random coin matrices of this sort below.

Consider  $\Omega=\T^{\cT_q\times A_q}$, $\T$ being the torus, as a probability space with $\sigma$ algebra generated by the cylinder sets and measure $\P=\otimes_{x\in\cT_q\atop \tau\in A_q}d\nu$ where $d\nu$ is a probability measure on $\T$. 
Let $\{\omega^\tau_x\}_{x\in \cT_q, \tau\in A_q }$ be a set of i.i.d. random variables on the torus $\T$ with common distribution $d\nu$. We will note $\Omega\ni \omega=\{\omega^\tau_x\}_{x\in \cT_q, \tau\in A_q }$ and we assume that
$
d\nu(\theta)=l(\theta)d\theta, 
$
with $l\in L^\infty(\T)$,
the support of which has non-empty interior.
We define a random diagonal unitary operator on $\cK_q$ by 
\be\label{dd}
\D_\omega x\otimes \tau = e^{i\omega^\tau_x} x\otimes \tau, \ \ \forall (x,\tau)\in \cT_q\times A_q.
\ee
The random version of our quantum walks is defined by the unitary operator
\be
U_\omega(C)=\D_\omega U(C) \ \ \mbox{on } \cK_q.
\ee
This  amounts to replacing the constant matrix $C\in \C^{q}$ by a family of site dependent random matrices $C_\omega(x)\in \C^{q}$, $x\in \cT_q$,  acting as in (\ref{explicitsitedep}) with $C_\omega(x)_{ab}=e^{i\omega^{a}_{xa}}C_{ab}$ ,  for $q$ even, and $C_{a_ja_k}(x_e)=e^{i\omega^{a_j}_{x_e a_{j+1}}}C_{a_ja_k}$ and
$C_{a_ja_k}(x_o)=e^{i\omega^{a_j}_{x_o a_{j+2}}}C_{a_ja_k}$,  for $q$ odd.

These operators are ergodic in the following sense:  
Let $z\in \cT_q$;
 we use the same symbol $T_z$ to denote the measure preserving map $T_z: \Omega\ra \Omega$ defined by $T_z\omega =\{\omega_{zx}^\tau\}_{x\in \cT_q, \tau \in A_q}$ and the unitary operator on $\cK_q$ given by $T_z x\otimes \tau = zx\otimes \tau$. One has  $T_z^{-1}=T_{z^{-1}}=T_z^*$ on $\cK_q$ and $T_z^* \D_\omega T_z= \D_{T_z\omega}$. Moreover, 
for any $z$ such that $|z|$ is even, and  any $q$ we have
\be\label{ergodic}
T_z^* U_\omega(C) T_z = U_{T_z\omega}(C)
\ee
and the same holds for any function of $U_\omega(C)$. Finally, the random unitary operator $U_\omega(C)$ on $\cK_q$  depends
continuously on the coin matrix $C$: For any coin matrices $C, C'\in U({q})$,
$
\|U_\omega(C)-U_\omega(C')\|_{\cK^q}=\|\I\otimes (C-C')\|_{\cK^q}=\|C-C'\|_{\C^{q}}.
$

\subsection{Spectral Criteria}

We shall repeatedly make use of the following general spectral criteria, see e.g. \cite{RS}.
Let $U$ be a unitary operator on a separable Hilbert space $\cH$. The spectral measure $d\mu_\phi$ on the torus $\T$ associated with a normalized vector $\phi\in \cH$ decomposes as
$
d\mu_\phi=d\mu^{pp}_\phi+d\mu^{ac}_\phi+d\mu^{sc}_\phi
$
 into its pure point, absolutely continuous and singular continuous components. The corresponding supplementary orthogonal spectral subspaces are denoted by $\cH^{\#}(U)$, with $\#\in\{pp, ac, sc\}$.
The Fourier coefficients of the spectral measure read
\be
\hat\mu_{\phi}(n)=\overline{\hat\mu_{\phi}(-n)}=\bra \phi | U^n \phi\ket=\int_\T e^{i\theta n}d\mu_{\phi}(\theta), \ \ \forall n\in\Z.
\ee
Then, Wiener or RAGE Theorem says that
\be
\lim_{N\ra\infty}\frac{1}{N}\sum_{n=0}^{N}|\bra \phi | U^n \phi\ket|^2=\sum_{\theta\in\T}(\mu^{pp}_\phi\{\theta\})^2,
\ee
whereas the absolutely continuous spectral subspace of $U$, $\cH^{ac}(U)$, is given by
\be\label{critac}
\cH^{ac}(U)
=\overline{\Big\{\phi \ | \ \sum_{n\in \N}|\bra \phi | U^n \phi\ket|^2<\infty \Big\}}.
\ee
Given $\{P_r\}_{r\in\N}$ a family of finite rank orthogonal projectors such that $\lim_{r\ra\infty }P_r=\I$ in the strong sense, one has the following.  The vector $\phi$ belongs to $\cH^{c}(U)=\cH^{ac}(U)\oplus\cH^{sc}(U)$, the continuous spectral subspace of $U$, if and only if for any $r\geq 0$
\be\label{pointcrit}
\lim_{N\ra\infty}\frac{1}{N}\sum_{n=0}^{N}\|P_rU^n\phi\|=0.
\ee
The vector $\phi$ belongs to the pure point spectral subspace of $U$, $\cH^{pp}(U)$, if and only if for any $r\geq 0$
\be
\lim_{r\ra\infty}\sup_{n\in \N}\|(\I-P_r)U^n\phi\|=0.
\ee

When  criterion (\ref{critac}) is
 applied to vectors from an orthonormal basis of $\cH$,  $\{e_j\}_{j\in I}$, $I$ a discrete  set of indices, one expands to get
 \be\label{defpath}
\bra e_{k_0} | U^n e_{k_n}  \ket=\sum_{(k_1, k_2, \dots, k_{n-1})\in I^{n-1}} 
\bra e_{k_0} | U e_{k_1}  \ket \bra e_{k_1} | U e_{k_2}  \ket\dots \bra e_{k_{n-1}} | U e_{k_{n}}  \ket,
\ee
and we consider each sequence $\{k_0, k_1, k_2, \dots, k_{n}\}\in I^{n+1}$ as a path with complex weight given by the product of matrix elements of $U$ in the summand above. If $U$ has a matrix representation with band structure, the sum (\ref{defpath}) is finite. 

Note that for $U(C)$ on $\cK_q$ given by (\ref{defeven}) and (\ref{defodd}), the diagonal elements of $U^{2n+1}(C)$ in the canonical orthonormal basis (\ref{canbas}) are all zero, for $n\in \N$. Moreover, 
in case $C=\I$, $\bra x\otimes \tau | S^{2n} x\otimes\tau \ket=\delta_{0,n}$, for all $x\otimes \tau \in \cK_q$. Hence $d\mu_{x\otimes\tau}=\frac{d\theta}{2\pi}$ and $\sigma(S)=\sigma_{ac}(S)=\U$.

\subsection{Permutation Matrices}

We consider here coin matrices  given by permutation matrices which lead to an explicit spectral analysis. We do not attempt to 
cover all cases 
but, for both cases $q$ odd and $q$ even, we 
analyze two permutations around which we shall perturb later on. 

Let $\pi\in {\mathfrak S}_q$ that we view as acting on $A_q$. Then $C_\pi=\sum_{\tau\in A_q}|\pi(\tau)\ket\bra\tau |$ is the corresponding permutation matrix. We will generalize this set of special matrices by allowing the matrix elements of $C_\pi$ to carry phases. We introduce 
$
\Phi=\mbox{diag }(e^{i\ffi_{a_1}}, e^{i\ffi_{a_2}}, \cdots, e^{i\ffi_{a_q}})\in U(q)
$
and $C_\pi^\Phi$ by
\be\label{cphi}
C_\pi^\Phi=\Phi C_\pi=\sum_{\tau\in A_q}e^{i\ffi_{\pi(\tau)}}|\pi(\tau)\ket\bra\tau |,
\ee
that we call a decorated permutation matrix.
Among all decorated permutation matrices, those $C_\pi^\Phi$ which give rise to pure point spectrum for $U_\omega(C_\pi^\Phi)$ with finite dimensional cyclic subspaces for all $\omega$ play a special role. Such coin matrices are called {\em fully localizing matrices} 
The set of fully localizing permutation matrices will be denoted by $\Lambda\in U(q)$. We exhibit an element of $\Lambda$ for any $q$ in the following lemma.

\begin{lem}\label{locqodd}
Let $q$ be odd and let $C^\Phi_{(12\cdots q)}$ be the decorated permutation matrix corresponding to $(12\cdots q)\in\mathfrak{S}_{q}$. Then $U_\omega(C^\Phi_{(12\cdots q)})$ has pure point spectrum and admits 
\bea\label{invxo}
\cH_{x_o}={\mbox{span }}&\{&  x_o\otimes a_{1}, x_oa_4\otimes a_{2}, x_o\otimes a_3, \cdots x_o\otimes a_{q-2}, x_oa_{1}\otimes a_{q-1}, x_o\otimes a_q \nonumber\\
&&x_oa_3\otimes a_1, x_o\otimes a_{2}, \cdots  x_oa_{q}\otimes a_{q-2}, x_o\otimes a_{q-1}, x_oa_2\otimes a_q\},
\eea
for any odd $x_o\in\cT_q$, as invariant subspaces.
Moreover, $\oplus_{x_o\in\cT_q} \cH_{x_o}=\cH$ and
\be
\sigma(U_\omega(C^\Phi_{(12\cdots q)})|_{\cH_{x_o}})=e^{i\theta^{x_o}_\omega/(2q)}e^{i\ffi}\{1, e^{i2\pi/2q}, \cdots, e^{i2\pi(2q-1)/2q}\},
\ee
where  $\ffi=\frac{1}{q}\sum_{j=1}^q{\ffi_{a_j}}$ and 
$
\theta^{x_o}_\omega=\sum_{j=1}^q(\omega^{a_{j+1}}_{x_oa_{j+3}}+\omega_{x_o}^{a_j}).
$

{For $q$ even, $U_\omega (C^\Phi_{(1\  q/2+1)(2\ q/2+2)\cdots(q/2\  q)})$ has pure point spectrum and admits
\bea\label{invxe}
\cH_{x_o}={\mbox{\em span }} \bigcup_{j\in \{1,\dots, q\}}&\{ x_o\otimes a_{j}, x_oa_{j+q/2}\otimes a_{j+q/2}\}\equiv \bigoplus_{j\in \{1,\dots, q\}}\cH_{x_o\otimes a_j},
\eea
 for any odd $x_o\in\cT_q$, as  invariant subspaces. Moreover, $\oplus_{x_o\in\cT_q} \cH_{x_o}=\cH$ and
\be
\sigma(U_\omega(C^\Phi_{(1\  q/2+1)(2\  q/2+2)\cdots(q/2\   q)})|_{\cH_{x_o\otimes a_j}})=e^{i\tilde{\theta}^{x_o,j}_\omega/2}e^{i\tilde{\ffi}_j}\{1, e^{i\pi}\},
\ee
where  $\tilde{\ffi}_j=\frac{1}{2}(\ffi_{a_j}+\ffi_{a_{q/2+j}})$ and $\tilde{\theta}^{x_o,j}_\omega= (\omega^{a_{j+q/2}}_{x_oa^{-1}_{j}}+\omega_{x_o}^{a_j})$.}

The random variables $\{{\theta}^{x_o}_\omega\}_{x_o\in\cT_q}$, respectively $\{\tilde{\theta}^{x_o,j}_\omega\}^{j\in \{1,\dots, q\}}_{x_o\in\cT_q}$, are i.i.d  and distributed according to the $2q$-fold convolution $d\nu*d\nu*\dots d\nu$, respectively according to $d\nu*d\nu$.
\end{lem}

\begin{rem}   There exist other  fully localizing coin matrices, see the analysis below of the cases $q=3$ and $q=4$, which have similar properties.
\end{rem}
{\bf Proof:} This is a deterministic result, so we assume without loss that $\Phi=\un$. Take $q$ odd, explicit computations show that the list of vectors in (\ref{invxo}) correspond to the successive images of any of them by $U(C_{(12\cdots q)})$, so that $U(C_{(12\cdots q)})^{2q}|_{\cH_{x_o}}=\I_{\cH_{x_o}}$. Adding phases via the diagonal operator $\D_\omega$ preserves invariance of $\cH_{x_o}$ and turns the previous identity into $U_\omega(C_{(12\cdots q)})^{2q}|_{\cH_{x_o}}=e^{\theta^{x_o}_\omega}\I_{\cH_{x_o}}$, from which we get the spectrum of this restriction. We conclude by observing that $\oplus_{x_o\in\cT_q} \cH_{x_o}=\cH$.  The case $q$ even is dealt with similarly.
\ep
\medskip 

Examples of permutation matrices which give rise to absolutely continuous spectrum for the corresponding random quantum walk include $C_{(1)(2)\cdots (q)}=\I$ for all $q$, and $C_{(12\cdots q)}$ for $q$ even, as criterion (\ref{critac}) shows.
We'll come back to these cases below.

\section{Strong Disorder Localization} \label{secloc}

We prove here localization of $U_\omega(C)$ in regimes where the coin matrix $C$ is close enough to 
a fully localizing permutation matrix, which defines the {\it strong disorder regime}.
The strategy 
we use on $\cT_q$ follows \cite{J3}, making use of the fractional moments method
\cite{AM} adapted to the unitary framework in \cite{HJS2}. As in the self-adjoint case, the fractional moments method carries over from $\Z^{d}$ to Cayley trees easily, see e.g. \cite{A}, so that we don't spell out the details. 
We first define finite volume restrictions of random quantum walks.

Making use of Lemma \ref{locqodd}, we define boundary conditions which preserve unitarity and restrain the motion of the walker to balls of finite volume on $\cT_q$.
Let $C\in U(q)$ be given, $\pi_o=(1 2 \dots q)$, for $q$ odd and  { $\pi_e=(1\  \frac{q+2}{2})(2\  \frac{q+4}{2})\cdots (\frac{q}{2}\ q)$}, for $q$ even. Note by $C^\Phi_{\tilde \pi}$,  the decorated permutation matrix associated with $\tilde \pi=\pi_o$ if $q$ is odd and $\tilde \pi=\pi_e$ if $q$ is even. 
Consider $x_o\in \cT_q$ an odd site and define a site-dependent family of matrices $\cC_{x_o}=\{C(x)\in U({q})\}_{x\in\cT_q}$ by
\be
C(x)=\left\{\begin{matrix} 
C^\Phi_{\tilde \pi} & \mbox{if} \ d(x,x_o)\leq 1 \cr
C\ \ & \mbox{otherwise.\ \ \  \ \ }
\end{matrix}\right.
\ee
Since $U(\cC_{x_o})$ acts as $U(C^\Phi_{\tilde \pi})$ in the neighborhood of $x_o$,  Lemma \ref{locqodd} implies
\begin{lem}\label{bc}
For $q$ odd, respectively $q$ even, $U(\cC_{x_o})$ given by (\ref{explicitsitedep}) admits  $\cH_{x_o}$ defined by (\ref{invxo}), respectively (\ref{invxe}), as $2q$-dimensional  invariant subspace. 
\end{lem}
\begin{rem}  The same result with the same proof holds for 
$
U_\omega(\cC_{x_o})=\D_\omega U(\cC_{x_o}).
$
\end{rem}

Using such boundary conditions, we define for all $q$
restrictions of $U_\omega(C)$ to finite dimensional subspaces associated with balls $\Lambda_L(x_e)\subset \cT_q$ of odd radius $L\in 2\N+1$,  centered at even sites $x_e\in \cT_q$:
Let $L\in 2\N+1$ and $x_e\in \cT_q$ be an even site and consider the site-dependent family of coin matrices $\cC_{L, x_e}=\{C(x)\in U({q})\}_{x\in\cT_q}$ defined by
\be
C(x)=\left\{\begin{matrix} 
C^\Phi_{\tilde{\pi}} & \mbox{if} \ d(x,x_e)\in\{L-1,L,L+1\} \cr
C\ \ & \mbox{otherwise,\hspace{3.3cm}}
\end{matrix}\right.
\ee
{where $\tilde{\pi}=\pi_0$ for odd $q$ and $\tilde{\pi}=\pi_e$ for even $q$.}
The following lemma holds, with the notations
\be\label{defbcl}
U^{L, x_e}(C)= U(\cC_{L, x_e}), \ \ \ U_{\omega}^{L, x_e}(C)=\D_\omega U^{L, x_e}(C), 
\ee
\begin{lem}\label{last} 
For any $\omega \in \Omega$, the subspaces
$
\cH_{\Lambda_L(x_e)}=\bigoplus_{ \mbox{\tiny odd }x_o \in \cT_q \atop d(x_o,x_e)\leq L}\cH_{x_o}
$ 
and $\cH_{\Lambda_L(x_e)}^\perp$ are invariant under $U_{\omega}^{L, x_e}(C)$. {For odd $x_o$, the subspaces $\cH_{x_o}$ are given by (\ref{invxo}), respectively (\ref{invxe}), for $q$ odd, respectively $q$ even.}
Moreover,  $\dim \cH_{\Lambda_L(x_e)}=\frac{2q}{q-2}((q-1)^{L+1}-1)$
 and $\|U_{\omega}^{L, x_e}(C)-U_\omega(C)\|\leq \|C-C^\Phi_{\tilde \pi}\|_{\C^q}$.
\end{lem}
\begin{rem} Finite volume restrictions of the same sort can be constructed using any fully localizing permutation matrix.
\end{rem}
{\bf Proof:} By Lemma  \ref{bc},  the subspace $\bigoplus_{\mbox{\tiny odd }x_o\in \cT_q \atop d(x_o,x_e)=L}\cH_{x_o}$ is invariant.  Since sites $x,y$ with $d(x,x_e)\leq L-1$ and $d(y,x_e)\geq L+1$ are at least a distance 2 apart from each other, $\bra x\otimes a_k | U_\omega (\cC_{L, x_e}) y\otimes a_k\ket=0$, for all $j,k\in \{1, \dots, q\}$, which shows that $\cH_{\Lambda_L(x_e)}$ is invariant. The dimension of $\cH_{\Lambda_L(x_e)}$ is determined by Lemmas \ref{bc} and by the number of sites $x$ such that $|x|=l$ which is $q(q-1)^{l-1}$. Summing over all odd $l$ up to $L$ gives the result. The last estimate is straightforward from (\ref{explicitsitedep}).\ep\\

The finite volume unitary operator associated to the ball $\Lambda_L(x_e)$ is defined as the restriction
\be\label{finvolres}
U^{\Lambda_L(x_e)}_\omega(C)=U_\omega^{L, x_e}|_{\cH_{\Lambda_L(x_e)}}
\ \ \mbox{and } \  \ \
U^{\Lambda^C_L(x_e)}_\omega(C)=U_\omega^{L, x_e}|_{\cH^\perp_{\Lambda_L(x_e)}}.
\ee
As in Lemma \ref{last}, we have for any $C, C'\in U(q)$,
\be\label{normdiff}
\|U^{\Lambda_L(x_e)}_\omega(C)-U^{\Lambda_L(x_e)}_\omega(C')\|\leq \|C-C'\|_{\C^q}.
\ee

The Green function of $U_\omega(C)$ is denoted by
\begin{equation}
G_{a_j,a_k,\omega}(x,y;C,{z})=\langle x\otimes a_j | \, (U_\omega(C)-z)^{-1}  y\otimes a_k\rangle
\end{equation} 
and the finite volume Green function is denoted by $G^{\Lambda_L(x_e)}_{a_j,a_k,\omega}(x,y;{z})$, with $U_\omega^{\Lambda_L(x_e)}(C)$ in place of $U_\omega(C)$. We estimate the fractional moments of the finite volume resolvent 
and we take the limit $L\ra\infty$ to get suitable estimates on the  fractional moments of the full resolvent. The behavior in $L$ of the size of the boundary of the ball of radius $L$  being exponential on $\cT_q$ rather than algebraic on the lattice, 
we need to prove that the fractional moment estimates have an arbitrarily large exponential decay.

We prove in Appendix the following fractional moments estimate on the tree.
\begin{thm} \label{fme} Let $\pi_o
\in {\mathfrak S}_{q}$ be such that $C^\Phi_{\pi_o}\in \Lambda\subset U(q)$.
For all $0<s<1/3$, and all $\gamma>0$, there exist $K(s, \gamma)<\infty$ and $\eps(s,\gamma) >0$ such that for all $C\in U(q)$ with $\|C-C^\Phi_{\pi_0}\|\leq \eps(s,\gamma)$,  all $x,y\in \cT_q$ with $d(x,y)>2$, all $z\not\in \U$, and all $j,k\in \{1,\dots, q\}$, 
\be\label{am}
\E(|G_{a_j,a_k,\omega}(x,y;C,{z})|^s)\leq K(s, \gamma)e^{-\gamma d(x,y)}.
\ee
The estimate also holds for $\gamma=0$, without restriction on $C$ or $d(x,y)$.
\end{thm}

\begin{cor}\label{cor1}
Under the hypotheses of Theorem \ref{fme}, and for $\gamma>0$ large enough, 
\be
\sigma(U_\omega(C))=\sigma_{pp}(U_\omega(C)) \ \mbox{almost surely.}
\ee 
\end{cor}
{\bf Proof: } The result follows from the criteria (\ref{pointcrit}) applied to all basis vectors with $P_r$ the projector on the span of $\{x\otimes a \, | \, a\in A_q, |x|\leq r\}$, along the lines of Proposition 3.1 in \cite{HJS2}. Taking the decay rate $\gamma$ large enough allows us to compensate for the exponential growth in $r$ of $\dim P_r\leq c(q-1)^r$ on trees. \ep

\section{Weak Disorder Delocalization} \label{secdeloc}

 In this section, we prove that on any tree $\cT_q$, there exists special 
permutation matrices such that the spectrum of $U_\omega(C)$ is purely absolutely continuous, provided $C$ is close enough to these permutation matrices. This defines the {\it weak disorder regime} in this framework. We call these special permutation matrices {\em fully delocalizing} and the set they form will be denoted by $\cS$.
Our delocalization result is in keeping with the one Klein proved for the Anderson model on trees, \cite{Kl}. However, our statement is stronger in the sense that it is deterministic, see Remark \ref{remdeloc}, whereas it is known that the Anderson model on trees with radially symmetric random potential displays for all coupling constant purely singular spectrum, almost surely, see \cite{ASW}.

{

\subsection{Delocalization close to $C=\Phi$ for $q$ odd} 

Let $\pi=\mbox{Id}=(1)(2)\dots (q)$ be the identity permutation in ${\mathfrak S}_q$ so that $C^\Phi_{\mbox{\tiny Id}}=\Phi$.
\begin{prop}\label{weakdelocodd} Let $q\geq 3$ be odd and  $\epsilon=1/(4q^2(q-1))$. Then,  for any $\Phi=\mbox{diag}(e^{i\ffi_{a_j}})\in U(q)$, $\|C-\Phi\| \leq \epsilon$ implies for any $\omega\in \Omega$ that 
$
\sigma(U_\omega(C))=\sigma_{ac}(U_\omega(C)).
$
\end{prop}
{\bf Proof:} 
We  consider $C=\Phi +E$, where $E\in M_q(\C)$ is such that $\|E\|\leq \epsilon$ and $\Phi+E\in U(q)$.
The argument consists in showing that there exist $C, \kappa >0$ such that for all $x\in \cT_q$, $\tau\in A_q$ 
\be\label{estone1}
|\bra x\otimes \tau | U^{2n}_\omega(C) x\otimes \tau\ket|\leq C(\kappa \epsilon)^n/n^{3/2}.
\ee
This implies that $x\otimes \tau \in \cH^{ac}(U_\omega(C))$ if
$\epsilon\leq1/\kappa$, according to (\ref{critac}).
We introduce $0<\gamma\leq 1$ such that $|e^{i\ffi_a}+E_{a,a}|\leq \gamma$, for all $a\in A_q$. Separating the part on $l^2(\cT_q)$ from that on $\C^q$ of the basis vectors $y\otimes \sigma$, each path contributing to (\ref{estone1}) in the decomposition (\ref{defpath}) has a trace on $\cT_q$ of the form
\be\label{tpath1}
xa_{i_1}a_{i_2}a_{i_3}a_{i_4}\dots a_{i_{2n}}, \ \mbox{where $a_{i_j}\in A_q$ and  $a_{i_1}a_{i_2},\dots a_{i_{2n}}=e$}.
\ee
The corresponding sequence of coin variables depends on the parity of $x$: 
\bea\label{tcoineven}
&&\tau a_{i_1-1}a_{i_2-2}a_{i_3-1}a_{i_4-2}\dots a_{i_{2n}-2}, \ \mbox{where $\tau, a_{i_j}\in A_q$ and  $a_{i_{2n}-2}=\tau$}, \mbox{ if $|x|$ even
}\\
\label{tcoinodd}
&&\tau a_{i_1-2}a_{i_2-1}a_{i_3-2}a_{i_4-1}\dots a_{i_{2n-1}}, \ \mbox{where $\tau, a_{i_j}\in A_q$ and  $a_{i_{2n}-1}=\tau$}, \ \mbox{if $|x|$ odd.}
\eea
The weight of these paths is bounded above in modulus by $\eps^{2n-j}\gamma^{j}$, for some $0\leq j\leq 2n$ counting the number of diagonal elements of $C$, see (\ref{explicitsitedep}). We show that $j\leq n$.
In the list of matrix elements that constitute the weight of the path, there are $k\geq 0$ sequences of consecutive diagonal elements of length $m_i$, $i=1, 2,\dots, k$ so that there are $r=2n-\sum_{i=1}^k m_i=2n-j$ off diagonal elements. Each of the $m_i$ diagonal elements correspond to a sequence of the form (\ref{tcoineven}) or (\ref{tcoinodd}) which form an irreducible word by definition. Moreover, different such sequences cannot reduce one another and they must be separated by elements associated to off-diagonal elements. Since the irreducible words can only be reduced by the $r$ letters corresponding to off diagonal elements, the total length of the reduced word made of $2n$ letters is bounded below by $\sum_{i=1}^k m_i-r=2(j-n)$. Hence the requirement $j\leq n$. 
Finally,  for any $q\geq 3$, $\cN_q(2n)$, the number of paths of length $2n$ from $x$ to $x$ in $\cT_q$, is given for large $n$ by
\be\label{woess}
\cN_q(2n)={\tilde C(q)}\frac{(4(q-1))^n}{n^{3/2}}\left(1+O({n^{-1/2}})\right),
\ee
for some finite constant ${\tilde C(q)}$, see e.g.\cite{W}.
Taking into account the $q$ coin variables at each step, the number of contributing paths of the form (\ref{tpath1}) is less than $C{\kappa^n}/{n^{3/2}}$, with $\kappa=4q^2(q-1)$, which proves (\ref{estone1}).
\ep

\subsection{Delocalization close to $C_{(12\cdots q)}^\Phi$ for $q$ even}

A very similar argument allows us to prove a delocalization result for $q>2$ even.
Consider the permutation $(12\cdots q)$ and the corresponding decorated permutation matrix $C_{(12\cdots q)}^\Phi$. This matrix gives rise to cyclic subspaces whose trace on $\cT_q$ can be viewed as spirals, see Figure \ref{realspiral} for the case $q=4$.

\begin{prop} \label{weakdeloceven}  Let  $q>2$ be even and $\epsilon=1/(4q^2(q-1))$.
Then, for any $\Phi=\mbox{diag}(e^{i\ffi_{a_j}})\in U(q)$, $\|C-C_{(12\cdots q)}^\Phi\|\leq\epsilon$ implies for any $\omega\in \Omega$ that 
$
\sigma(U_\omega(C))=\sigma_{ac}(U_\omega(C)).
$
\end{prop}
\begin{rem} \label{remdeloc}
Both Propositions \ref{weakdelocodd} and \ref{weakdeloceven} are deterministic results which extend to families of unitary matrices $\cC=\{C(x)\}_{x\in \cT_q}$ of the sort considered in (\ref{explicitsitedep}), provided $C(x)$ satisfies the hypothesis for each $x\in\cT_q$. 
\end{rem}

\section{Spectral Diagrams for $q=3$ and $q=4$}\label{q=3q=4}

Focusing on the spectral diagrams for $q=\{3,4\}$, we exhibit families of coin matrices for which $U_\omega(C)$ has pure point spectrum almost surely, purely absolutely continuous spectrum for any $\omega$ or has mixed spectrum. These families allow us to describe the spectral transition.

\subsection{Permutation Coin Matrices for $q=3$}

For the case $q=3$, we sketch a more complete spectral diagram in Section \ref{sdq3}. The corresponding picture is given in Figure \ref{phase}.

The alphabet is denoted by $A_3=\{a,b,c\}$ and the orthonormal basis of the coin Hilbert space is denoted by $\{|a\ket, |b\ket,|c\ket\}$. The coin dependent shift $S$ on $\cK_3=\cT_3\otimes\C^3$ then reads
\be
S=S_{bc}\otimes |a\ket\bra a|+S_{ca}\otimes |b\ket\bra b|+S_{ab}\otimes |c\ket\bra c|
\ee
and all  coin matrices $C$ are written as $3\times 3$ matrices in the basis ordered as above. 
We refrain from decorating the permutation matrices by phases $\Phi$ in this section. We shall simply comment wherever necessary  on the modifications  required to generalize the statement made to the case of decorated permutation matrices.

The six different permutations of $\{a,b,c\}$ give rise to coin matrices inducing walks $U_\omega(C)$ with the following spectral properties, for any deterministic choice of diagonal $\D_\omega$: 
\begin{itemize}
\item  
$
\Lambda=
\{C_{(abc)}, C_{(acb)}\}
$,
 is the set of fully localizing matrices,
\item $\cS=\{C_{(a)(b)(c)}\}$ is the set of fully delocalizing matrix.
\item  $\cM=\{
C_{(a)(bc)}
, C_{(b)(ac)}
, C_{(c)(ab)}
\}$ 
 give rise to quantum walks with mixed spectra. 
\end{itemize}

To show the last point, we consider the first matrix of the list only. 
The operator $U_\omega(C_{(a)(bc)})$ leaves the subspace $l^2(\cT_3) \otimes |a\ket $ invariant, which gives rise to a shift essentially driven by $S_{bc}$ on the corresponding cyclic subspaces 
$\overline{\mbox{span }}\{\dots x_ecb\otimes a, x_ec\otimes a, x_e\otimes a, x_eb\otimes a, x_ebc\otimes a, \dots \}$ labelled by $x_e\in \cT_3$. Moreover, $U_\omega(C_{(a)(bc)})$ gives rise to another shift on the cyclic subspaces 
$\overline{\mbox{span }}\{\dots x_ebc\otimes c, x_eb\otimes b, x_e\otimes c, x_ec\otimes b, x_ecb\otimes c, \dots \}$ labelled by $x_e\in \cT_3$ with alternating coin state, essentially driven this time by $S_{cb}=S_{bc}^*$.
Finally, for all $x_e\in\cT_3$, the two-dimensional subspace $\overline{\mbox{span }}\{x_e\otimes b, x_ea\otimes c\}$ is invariant under $U_\omega(C_{(a)(bc)})$. Therefore, 
\be
\sigma(U_\omega(C_{(a)(bc)}))=\sigma_{pp}(U_\omega(C_{(a)(bc)}))\cup \sigma_{ac}(U_\omega(C_{(a)(bc)}))=\U.
\ee

\subsection{Delocalizing and Localizing Matrices for $q=3$}

We introduce here three one-parameter families of coin matrices $\{C^d_j(r)\}^{j\in\{1,2,3\}}_{0<r<1}$ which give rise to absolutely continuous operators $U_\omega(C^d_j(r))$, for any choice of phases $\D_\omega$.

For $0\leq r\leq 1$ and $t=\sqrt{1-r^2}$, set
\be
C^d_1(r)=\begin{pmatrix} 1& 0& 0 \cr 0&r&t \cr 0 & -t& r  \end{pmatrix}, C^d_2(r)=\begin{pmatrix} r& 0& t \cr 0&1&0\cr -t & 0& r  \end{pmatrix}, C^d_3(r)=\begin{pmatrix} r& t& 0 \cr -t&r&0\cr 0 & 0& 1  \end{pmatrix}.
\ee
If $r=1$, all matrices reduce to $\I=C_{(a)(b)(c)}$ and for $r=0$ they are correspond up to phases to the permutation matrices $C_{(a)(bc)}, C_{(b)(ac)}, C_{(c)(ab)}$.
\begin{lem}
For any  $0<r\leq 1$, $j\in\{1,2,3\}$ and any deterministic choice of $\D_\omega$  
\be
\sigma(U_\omega(C^d_j(r)))=\sigma_{ac}(U_\omega(C^d_j(r))) =\U.
\ee
\end{lem}
{\bf Proof:} The case $r=1$ corresponds to the identity. 
We consider $U_\omega(C^d_2(r))$, the other cases being similar. When restricted to the invariant subspace $l^2(\cT_3)\otimes |b\ket$ 
this operator gives rise to a shift which has absolutely continuous spectrum $\U$.  Consider the restriction to the coin variables $|a\ket, | c\ket$. For $0<r<1$, 
$C^d_2(r)$ makes the walker jump on $\cT_3$ from $x_e$ to $x_ea$ and $x_eb$ and from $x_o$ to $x_ob$ and $x_oc$ only. Therefore, as soon as a path contains a step $x_ea$ or $x_oc$, it is impossible to get back to $x_e$ or $x_o$. Thus, for any $x_e, x_o\in\cT_3$ and any $n\in\N$
\bea
|\bra x_e\otimes c | U_\omega^{2n}(C^d_2(r))  x_e\otimes c  \ket |=|\bra x_o\otimes a | U_\omega^{2n}(C^d_2(r))  x_o\otimes a  \ket |=t^{2|n|},
\eea
whereas all corresponding scalar products with other basis vectors yield $\delta_{0n}$. Since $t<1$, criterion (\ref{critac}) yields the result. \ep

\begin{rem}\label{remxxx}
The results holds for arbitrary site dependent alterations of the matrix elements  of $C^d_j(r)$ by phases which preserve unitarity. 
 Note also that different values of the parameter $0<r(x)\leq 1$ at different sites $x\in\cT_3$ are  allowed provided $\inf_{x}r(x)\geq r_0>0$. 
\end{rem}

We define here six other families of one-parameter coin matrices $\{C^l_j(r)\}^{j\in\{1,2, \dots,6\}}_{0<r<1}$ which give rise, almost surely, to pure point spectrum for the random operators $U_\omega(C^l_j(r))$. 

Consider for $0\leq r\leq 1$ and $t=\sqrt{1-r^2}$, 
\bea
&&C^l_1(r)=\begin{pmatrix} 0& r& t \cr 1&0&0\cr 0&-t & r  \end{pmatrix}, C^l_2(r)=\begin{pmatrix} 0& 1& 0 \cr r&0&t\cr -t&0 & r  \end{pmatrix}, C^l_3(r)=\begin{pmatrix} 0& 0& 1 \cr -t&r&0\cr r& t & 0  \end{pmatrix}, \nonumber\\
&&C^l_4(r)=\begin{pmatrix} 0& t& r\cr 0&r&-t \cr 1&0 & 0  \end{pmatrix}, C^l_5(r)=\begin{pmatrix}  r&0& -t \cr t&0&r\cr 0&1 & 0  \end{pmatrix}, C^l_6(r)=\begin{pmatrix} r& -t& 0 \cr 0&0&1\cr t&r & 0  \end{pmatrix}.
\eea
Note that for $r=1$ these matrices reduce by pairs to one of the permutation matrices $C_{(a)(bc)}, C_{(b)(ac)}, C_{(c)(ab)}$, whereas for $r=0$, they are correspond, up to phases, to the permutation matrices $C_{(abc)}$, respectively  $C_{(acb)}$, for odd, respectively even indices.

\begin{prop}\label{loc3}
For all  $0< r< 1$, and $j\in\{1,2,\dots,6\}$ we have almost surely 
\be
\sigma(U_\omega(C^l_j(r)))=\sigma_{pp}(U_\omega(C^l_j(r))).
\ee
\end{prop}
\begin{rem}
The same result holds if $C^l_j(r)\mapsto \Phi C^l_j(r)$, where $\Phi =\mbox{diag }(e^{i\ffi_a}, e^{i\ffi_b}, e^{i\ffi_c})$, where $\ffi_\#$ can possibly depend on $r$.
\end{rem}
{\bf Proof:}
Without loss, we can consider  the matrix $C^l_1(r)$ only.
The strategy is as follows. The shape of the matrices $C^l_j(r)$ is such that the one step evolution operator $U_\omega(C^l_j(r))$ admits cyclic subspaces in each of which it acts as a one-dimensional random unitary operator. Then transfer matrix methods allows us to prove localization for all values of $0<r<1$. 
We first determine the cyclic subspaces of $U_\omega(C^l_1(r))$.
\begin{lem}
The $U_\omega(C^l_1(r))$-cyclic subspaces $\cH_{x_e\otimes a}$ generated by the vectors $x_e\otimes a$, $x_e\in\cT_3$ an even site, are given by
\bea\label{cyclic1d3}
\cH_{x_e\otimes a}=\overline{\mbox{span }}&\{&\dots, x_eca \otimes b, x_eca \otimes c, x_e\otimes a, x_ec \otimes a, x_ec \otimes b, x_ec \otimes c,  \\ \nonumber 
&& x_ecb \otimes b, x_ecb \otimes c, x_ecbac \otimes a, x_ecba \otimes a,  x_ecba \otimes b,  x_ecba \otimes c , \cdots\}.
\eea 
Their direct sum over $x_e$ spans $\cK_3$, taking into account the identities 
\be \cH_{x_e\otimes a}=\cH_{x_ecabc\otimes a}=\cH_{x_ecbac\otimes a}, \ \ \forall x_e\in \cT_3.
\ee
\end{lem}
\begin{rem}
Graphically, the sites of $\cT_3$ involved in (\ref{cyclic1d3}) are depicted in figure \ref{fig1d3}. 
\end{rem}
\begin{figure}[htbp]
   \begin{center}
      \includegraphics[scale=.25, angle=-90]{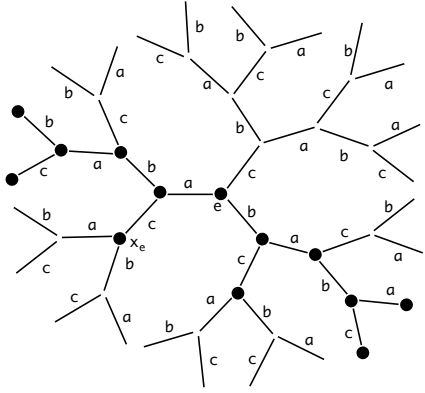}
   \end{center}
   \caption{\footnotesize Sites from the cyclic subspace $\cH_{x_e\otimes a}$.}\label{fig1d3}
\end{figure}

{\bf Proof:} One looks at the effect of powers of $U_\omega(C^l_1(r))$ on vectors related to the even site $y_e\in \cT_3$:
 First note that 
$|\bra y_o\otimes \tau| U_\omega(C^l_1(r)) y_e\otimes a\ket |=\delta_{y_o, y_ec}\delta_{\tau, b}$, which means that $y_e\otimes a$ is sent to $y_ec\otimes b$ by $U_\omega(C^l_1(r))$. On the other hand, $\bra y_e\otimes a | U_\omega(C^l_1(r)) y_o\otimes \tau \ket$ equals zero, unless $y_o=y_ec$ and $\tau\in\{b,c\}$. Hence the vector $y_e\otimes a$ is never connected to  $y_ea\otimes\tau$ or $y_eb\otimes\tau$, for any $\tau\in A_3$.  Similarly, if $\tau\in\{b,c\}$ ,$\bra y_ec\otimes \sigma | U_\omega(C^l_1(r)) y_e\otimes \tau \ket=0$, for all $\sigma\in A_3$, and 
the same is true for $\bra y_ec\otimes \sigma | U^*_\omega(C^l_1(r)) y_e\otimes \tau \ket$. In other words, the vectors $y_e\otimes \tau$, with $\tau\in\{b,c\}$, are never connected to $y_ec\otimes \sigma$, for any $\sigma\in A_3$. This is enough to reach the first conclusion of the lemma, while the second conclusion follows immediately. \ep

Consider now $U_\omega(C^l_1(r))|_{\cH_{x_e\otimes a}}$. While the order provided in (\ref{cyclic1d3}) allows for an easier identification of the periodicity, we use the following order to get a matrix representation of this operator: 
\bea\label{relabel}
&&\{\dots,  x_eca \otimes c, x_eca \otimes b, x_e\otimes a, x_ec \otimes c, x_ec \otimes a, x_ec \otimes b,  x_ecb \otimes c, x_ecb \otimes b, \nonumber\\
&& \hspace{4,0cm} x_ecbac \otimes a, x_ecba \otimes c,  x_ecba \otimes a,  x_ecba \otimes b , \cdots\}
\eea
We  denote these vectors by $e_j$, $j\in\Z$, in such a way that the set (\ref{relabel}) corresponds to
\bea
&&\{\dots,  e_{-1}, e_{0}, e_{1}, e_{2}, e_{3}, e_{4}, e_{5}, e_{6}, 
e_{7}, e_{8}, e_{9}, e_{10}, \cdots\},
\eea
and the cyclic subspace is equivalent to $l^2(\Z)$.
Therefore, $U(C^l_1(r))|_{\cH_{x_e\otimes a}}$ is equivalent to a seven-diagonal unitary matrix $V(C)$ in this basis 
\be\label{7diag}
U(C^l_1(r))|_{\cH_{x_e\otimes a}}\simeq V(C)=\begin{pmatrix}
\ddots & 0 &    &    &     &     &      &   &   &     \cr
              0      & 0  &   &     &  1 &      &   &   &   &        \cr
            0        &  0 & 0 & t  &  0 & r    &   &   &   &        \cr
            r         & -t  &  0& 0 & 0 &  0   &   &   &   &        \cr
                      &     & 0 & 0  &0 & 0    &  t & r&   &       \cr
                      &     & 1 &  0 & 0& 0    & 0  & 0 &   &       \cr
                      &     &   & r    &   &  -t  & 0  & 0 &   &       \cr
                      &     &   &      &   &       & 0  &  0&   &        \cr
                      &     &   &      &   &       &  0 &  0& 0 &  t     \cr
                      &     &   &     &   &        &  r & -t & 0  &     \ddots
\end{pmatrix},
\ee
where the  dots mark the diagonal, the first listed column is the image of $e_{-1}$ and the periodicity is six along the diagonal. The matrix representation of $U_\omega(C^l_1(r))|_{\cH_{x_e\otimes a}}$, denoted by $V_\omega(C)$, is obtained from (\ref{7diag}) by multiplying the row labelled by $j\in\Z$ by a phase $e^{i\omega_j}$, where $\{\omega_j\}_{j\in\Z}$ are distributed according to $d\nu$.
The spectral analysis of the above one-dimensional unitary random operator on $l^2(\Z)$ with a band structure can be performed by considering its generalized eigenvectors defined by $\psi=\sum_{j\in\Z}\psi(j)e_j$ and 
\be\label{gev}
(V_\omega(C)-z\I)\psi=0 \ \ \ \mbox{in $l^2(\Z)$,  where $z\in {\mathbb U}$}.
\ee
Explicit computations yield the transfer matrices below
\begin{prop}
For any $z\in \C^*$, the solutions of (\ref{gev}) are entirely determined by the sequence $\{(\psi(6j-1),\psi(6j))\}_{j\in \Z}$. Moreover, we have the relation
\be
\begin{pmatrix} \psi(6j+5) \cr \psi(6j+6))
\end{pmatrix}=T_z(j)\begin{pmatrix} \psi(6j-1) \cr \psi(6j)\end{pmatrix} \ \ \ \mbox{for any $j\in \Z$}
\ee
for 
$
T_z(j)=T_z(\omega_{6j}+\omega_{6j+3}, \omega_{6j+1}+\omega_{6j+4}, \omega_{6j+2}+\omega_{6j+5})\in M_2(\C)
$
given by
\be T_z(\alpha, \beta, \gamma)= \frac{(rz^2e^{-i\beta}-1)e^{i\gamma}}{rz^2(z^2e^{-i\beta}-r)}\begin{pmatrix}
r^2 & -tr \cr
-tr& (\frac{e^{-i(\alpha+\gamma)}z^4(z^2e^{-i\beta}-r)}{(rz^2e^{-i\beta}-1)}+t^2)\end{pmatrix}.
\ee
\end{prop}
\begin{rems}
i) If $z=e^{-i\lambda}\in{\mathbb U}$ we have $T_{e^{-i\lambda}}(\alpha, \beta, \gamma)=T_1(\alpha+2\lambda, \beta+2\lambda, \gamma+2\lambda)$ and
$
\det(T_{1}(\alpha, \beta, \gamma))=\left(\frac{re^{-i\beta}-1}{e^{-i\beta}-r}\right)e^{i(\gamma-\alpha)}\in {\mathbb U}.
$\\
ii) The expressions for the remaining coefficients read
\bea
&&\psi(6j+1)=     \frac{te^{i(\omega_{6j+2}-\omega_{6j+4})}(r\psi(6j-1)-t\psi(6j))}{(z^2e^{-i(\omega_{6j+1}+\omega_{6j+4})}-r)}, \  \psi(6j+3)=  ze^{-i\omega_{6j}}\psi(6j)     \\
&&\psi(6j+2)=   \frac{e^{i\omega_{6j+2}}}{z}(r\psi(6j-1)-t\psi(6j)) , \    \psi(6j+4)=   \frac{te^{i\omega_{6j+2}}(r\psi(6j-1)-t\psi(6j))}{z(z^2e^{-i(\omega_{6j+1}+\omega_{6j+4})}-r)} .     \nonumber 
\eea
iii) Replacing $C_1^l(r)$ by $\Phi C_1^l(r)$ amounts to shifting the random variables $\omega_j$ according to
\bea
&&\omega_{6j}\mapsto \omega_{6j}+\ffi_b, \hspace{.9cm}
\omega_{6j+1}\mapsto \omega_{6j+1}+\ffi_a, \hspace{.2cm}
\omega_{6j+2}\mapsto \omega_{6j+2}+\ffi_c, \nonumber\\
&&\omega_{6j+3}\mapsto \omega_{6j+3}+\ffi_a, \hspace{.2cm}
\omega_{6j+4}\mapsto \omega_{6j+4}+\ffi_b, \hspace{.2cm}
\omega_{6j+5}\mapsto \omega_{6j+5}+\ffi_c, 
\eea
for all $j\in\Z$. Consequently, this amounts to replace to the transfer  matrix  $T_z(\alpha, \beta, \gamma)$ by $\tilde T_z(\alpha, \beta, \gamma)=T_z(\alpha+\ffi_a+\ffi_b, \beta+\ffi_a+\ffi_b, \gamma+2\ffi_c)$.
\end{rems}

The random transfer matrices $\{T_z(j)\}_{j\in\Z}$ are i.i.d. so that we can follow 
\cite{BHJ}, \cite{HJS1} to prove spectral localization, via Shnol's and F\"urstenberg's Theorems. Since $d\nu$ is absolutely continuous 
with support of non empty interior, 
one needs to show that the group $\cG$ generated by products of transfer matrices 
is non compact and irreducible in an appropriate sense, in order to get a positive Lyapunov exponent. Concerning the first point we have 
\begin{lem}
Assume that $0<r<1$, and that there exists $\theta_0\neq\theta_1\in \T$ in the support of $d\nu$. Then $\cG$ is non compact.
\end{lem}
{\bf Proof:} We first get rid of the 
spectral parameter $z\in\C^*$ by making use of the following identities obtained by explicit computations. For any $z\in \C^*$ and any $0<r<1$
\bea\nonumber
T^{-1}_z(\alpha, \beta, \gamma)T_z(a,\beta,c)&=&\begin{pmatrix}
e^{i(c-\gamma)}&\frac{t}{r}(e^{i(\alpha-a)}-e^{i(c-\gamma)})\cr
0&e^{i(\alpha-a)}
\end{pmatrix}\equiv R(c-\gamma,\alpha-a),\\
T_z(a,\beta,c)T^{-1}_z(\alpha, \beta, \gamma)&=&
\begin{pmatrix}
e^{i(c-\gamma)}&0\cr
\frac{t}{r}(e^{i(\alpha-a)}-e^{i(c-\gamma)})&e^{i(\alpha-a)}
\end{pmatrix}\equiv L(c-\gamma, \alpha-a).
\eea
\begin{rem}
The maps $R$ and $L$ are invariant under the replacement of $C_1^l(r)$ by $\Phi C_1^l(r)$.
\end{rem}
Both maps $(\theta,\eta)\mapsto R(\theta,\eta)$ and $(\theta,\eta)\mapsto L(\theta,\eta)$ are group isomorphisms and we have
\be
L(\theta,\eta)=R^T(\theta,\eta), \ \ \ R(-\theta, -\eta)=\overline{R}(\theta,\eta) \ \ \ \Rightarrow R(-\theta, -\eta)=L^*(\theta,\eta).
\ee
We compute
\be
L(\theta,\eta)R(\alpha, \beta)=\begin{pmatrix}
e^{i(\theta+\alpha)} & \frac{t}{r}e^{i\theta}(e^{i\beta}-e^{i\alpha}) \cr
\frac{t}{r}e^{i\alpha}(e^{i\eta}-e^{i\theta})  & e^{i(\eta+\beta)}+ \frac{t^2}{r^2}(e^{i\beta}-e^{i\alpha})(e^{i\eta}-e^{i\theta}) 
\end{pmatrix}
\ee
s.t. $L(\theta,\eta)R(-\theta,-\eta)>0$, has determinant one and
\be
 \mbox{tr}(L(\theta,\eta)R(-\theta,-\eta))=
 2\left(1+\frac{t^2}{r^2}(1-\cos(\theta-\eta))\right).
\ee
Consequently, one eigenvalue of this matrix has modulus larger than one, if $\theta\neq\eta$ on $\T$. 
\ep

Concerning the second point, we introduce  $\tau: M_2(\C)\ra M_4(\R)$ defined by \be
\begin{pmatrix}a & b \cr c & d \end{pmatrix}\rightarrow
  \begin{pmatrix}\Re(a)I + \Im(a)J & \Re(b)I +
    \Im(b)J \cr \Re(c)I + \Im(c)J & \Re(d)I + \Im(d)J  \end{pmatrix},
\mbox{where} \
I = \left( \begin{array}{cc} 1 & 0 \\ 0 & 1
\end{array} \right), \ J = \left(
\begin{array}{cc} 0 & 1 \\ -1 & 0 \end{array} \right).
\ee This map is a homeomorphism from  $M_2(\C)$ to $\tau(
M_2(\C))$ and a group homeomorphisms from the set
of matrices in $M_2(\C)$ with determinant of modulus one to the
set of matrices in $M_4(\R)$ with determinant of modulus one. Irreducibility is expressed as follows.
\begin{lem} 
The set $\{\tau(T_{e^{-i\lambda}}(\alpha, \beta, \gamma))\in M_4(\R), (\alpha, \beta, \gamma)\in \mbox{supp }d\nu+\mbox{supp }d\nu\}$ is irreducible in $\R^4$ if the support of $d\nu$ has non empty interior.
\end{lem}
{\bf Proof:} It is enough to consider $T_1(\alpha, \beta, \gamma)$ with $(\alpha, \beta, \gamma)\in I^3$, where $I\subset \T$ is an arbitrary open arc. Then, with 
$
\frac{re^{-i\beta}-1}{e^{-i\beta}-r}=e^{-i\chi(\beta)}, \ \ A=\gamma-\chi(\beta), \ \ B=-\alpha
$,
we can write
\bea
\label{eqdiag1}
\tau(T_1(\alpha, \beta, \gamma))&=&\cos(A)M_1+\sin(A)M_2+\cos(B)N_1+\sin(B)N_2
\eea
with
\be
M_1=\begin{pmatrix}r &0 &-t &0 \cr
0 &r &0 &-t \cr
-t &0 &t^2/r & 0\cr
 0&-t &0 &t^2/r \end{pmatrix}, 
 M_2=\begin{pmatrix}0 &r &0& -t  \cr
-r &0 &t&0 \cr
0 &-t &0 & t^2/r\cr
 t&0 &-t^2/r &0 \end{pmatrix},
\ee
and
\be
N_1=\begin{pmatrix}0 &0 &0 &0 \cr
0 &0 & 0&0 \cr
0 &0 &1/r&0 \cr
 0& 0 &0 &1/r \end{pmatrix}, N_2=\begin{pmatrix}0 &0 &0 &0 \cr
0 &0 &0&0 \cr
0 &0 &0&1/r \cr
 0& 0 & -1/r&0 \end{pmatrix}.
\ee 
Keeping $\beta$ fixed, 
any nontrivial subspace $V\subset \R^4$ 
invariant under $\tau(T_1(\alpha, \beta, \gamma))$ has to be invariant under $M_1$, $M_2$, $N_1$ and $N_2$, since $A$ and $B$ are independent.  
Since these last two matrices are real (anti) self-adjoint, they leave
$V^\perp$ invariant as well. Hence, $V$ and $V^\perp$ are
generated by real eigenvectors of these matrices, if they are
diagonalizable over $\R$. 
If $\dim V=2$, it can be generated by $\{e_1,e_2\}$ or $\{e_3, e_4\}$ only, 
where $\{e_j\}_{j=1,\dots,4}$ is the canonical basis of $\R^4$. 
This is ruled out since these subspaces are not invariant under $M_1$.
Also, if $\dim V=1$, the only possibility is $V\subset {\mbox{span }}\{e_1, e_2\}$. The same argument forbids this and since it applies to $V^\perp$ as well, it takes care of the case where $\dim V=3$. 
\ep
\begin{rem}
If  $C_1^l(r)$ is replaced by $\Phi C_1^l(r)$, the same argument proves the Lemma since $\beta$ is fixed and $A$ and $B$ are given by $\gamma$ and $-\alpha$ plus a constant term in that case. 
\end{rem}
The arguments of \cite{HJS1} prove that Proposition \ref{loc3} derives from these properties. \ep

\subsection{Spectral Transition for $q=3$}\label{sdq3}
\begin{figure}[!h]
   \begin{center}
      \includegraphics[scale=.42]{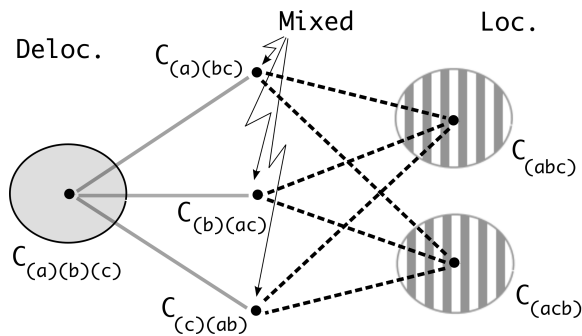}
   \end{center}
   \caption{\footnotesize Partial spectral diagram for $q=3$.}\label{phase}
\end{figure}

We showed the existence of six continuous paths in $U(3)$ from a small neighborhood of the set $\Lambda$ of fully localizing coin matrices to a small neighborhood of the set $\cS$ of  delocalizing coin matrices, through elements of the set $\cM$ of coin matrices inducing mixed spectra. 
Each element of $\Lambda$ is linked to an element of $\cM$ by means of the family $C^l_j(r)$, with suitable decorating phase $\Phi(r)$, on which almost sure localization takes place. And each element of $\cM$ is linked to the only element of $\cS$ by a path of the form $C^d_j(r)$, with suitable decorating phase $\Phi$, which induces absolutely continuous spectrum for all $\omega$ for the corresponding walk. The spectral diagram in Figure \ref{phase} doesn't show it explicitly, but as mentioned above, it holds for matrices decorated by phases $\Phi$ as well.

\subsection{Propagating, Reducing and Localizing Families for $q=4$}

We now turn to the case $q=4$.
Using the notations above, we describe a spectral transition for $q=4$ from $\Lambda$ to $\cS$ which is different from the case $q=3$ in the sense that it avoids elements from $\cM$.

The alphabet is denoted by $A_4=\{a,b,a^{-1},b^{-1}\}$ and the ordered orthonormal basis of the coin Hilbert space is denoted by $\{|a\ket, |b\ket,|a^{-1}\ket,|b^{-1}\ket\}$. 
The sites of the tree $\cT_4$ are labeled according to  Figure \ref{tree4} and
all $4\times 4$ coin matrices $C$ are written in the basis above.

We define the set of {\it propagating coin matrices} $\cP$, respectively of {\it reducing coin matrices} $\cR$ as unitary matrices of the form
\be
\begin{pmatrix}
* & * & 0 & * \cr
* & * & * & 0 \cr
0 & * & * & * \cr
* & 0 & * & * \cr
\end{pmatrix}\in \cP, \ \ \mbox{respectively} 
\begin{pmatrix}
* & 0 & * & 0 \cr
0 & * & 0 & * \cr
* & 0 & * & 0 \cr
0 & * & 0 & * \cr
\end{pmatrix}\in \cR.
\ee 
Propagating matrices $C^P\in\cP$ induce purely absolutely continuous spectrum for $U_\omega(C^P)$ since, in the language of (\ref{defpath}), it is impossible to come back to a site of $\cT_4$ already visited, irrespectively of the coin variable. Hence criteria (\ref{critac}) applies to all basis vectors and shows this is a deterministic property. This differs from Proposition \ref{weakdeloceven}, in that it is a non perturbative statement.

Reducing matrices $C^R\in\cR$ with non zero off diagonal elements induce pure point spectrum for $U_\omega(C^R)$, almost surely. Indeed, since they decouple the coin subspaces $\mbox{span}\{|a\ket,|a^{-1}\ket\}$ and $\mbox{span}\{|b\ket,|b^{-1}\ket\}$,
they reduce the analysis of $U_\omega(C^{R})$ to a direct sum of one dimensional 
random quantum walks taking place  on $l^2(\Z)\otimes \C^2$. Such walks give rise to dynamical localization almost surely, whatever the underlying deterministic coin matrix is, except in the diagonal case, see \cite{JM}.

In particular,  for the two parameter family 
\be\label{defred}
[0,2\pi]^2\ni(\psi,\xi)\mapsto  C^R(\psi,\xi)=\begin{pmatrix}
\cos(\psi) & 0 & \sin(\psi) & 0 \cr
0 & \cos(\xi) & 0 & \sin(\xi) \cr
-\sin(\psi) & 0 & \cos(\psi) & 0 \cr
0 & -\sin(\xi) & 0 & \cos(\xi) \cr
\end{pmatrix}
\ee
it holds 
\be\label{rpp}
 \ \mbox{ $\sigma(U_\omega(C^R(\psi,\xi)))=\sigma_{pp}(U_\omega(C^R(\psi,\xi)))$ a.s. $\Leftrightarrow \sin(\psi)\sin(\xi)\neq 0$.}
\ee
On the other hand,  for the families $[0,2\pi)^2\ni(\psi,\xi)\mapsto C^P_j(\psi,\xi)\in U(4)$, $j\in\{1,2,3\}$
\bea\label{fprop}
&&C^P_1(\psi,\xi)=\begin{pmatrix}
\cos(\psi) & 0 & 0 & \sin(\psi) \cr
0 & \cos(\xi) & \sin(\xi) & 0 \cr
0 & -\sin(\xi) & \cos(\xi) & 0 \cr
-\sin(\psi) & 0 & 0 & \cos(\psi) \cr
\end{pmatrix},\nonumber\\
&&C^P_2(\psi,\xi))=\begin{pmatrix}
0& \cos(\xi) & 0 & \sin(\xi) \cr
\cos(\psi) & 0 & \sin(\psi) & 0 \cr
0 & -\sin(\xi) & 0 & \cos(\xi) \cr
-\sin(\psi) & 0 & \cos(\psi) & 0 \cr
\end{pmatrix}, 
\nonumber\\
&&C^P_3(\psi,\xi)=\begin{pmatrix}
\cos(\psi) & \sin(\psi) & 0 & 0 \cr
-\sin(\psi) & \cos(\psi) & 0 & 0 \cr
0 & 0 & \cos(\xi) & \sin(\xi) \cr
0 & 0 & -\sin(\xi) &\cos(\xi) \cr
\end{pmatrix},
\eea
it holds for any realization $\omega$, any $(\psi,\xi)\in [0,2\pi)^2$ and any $j\in\{1,2,3\}$,
\be\label{pac}
\sigma(U_\omega(C^P_j(\psi,\xi)))=\sigma_{ac}(U_\omega(C^P_j(\psi,\xi))). 
\ee
Moreover, we have existence of localizing families of coin matrices:
\begin{lem}\label{1ppp}
The four one parameter families $[0,2\pi)\ni \psi\mapsto C_j(\psi)$, $j\in\{1,2,3,4\},$ 
\bea
&&C_1(\psi)=\begin{pmatrix}
 0 & 0& \cos(\psi) & \sin(\psi)  \cr
 0 & 0 & -\sin(\psi) & \cos(\psi) \cr
1 & 0 & 0 & 0 \cr
0 & 1 & 0 & 0 \cr
\end{pmatrix},
C_2(\psi)=\begin{pmatrix}
 0 & 0&  1 & 0 \cr
 0 & 0 &0 & 1 \cr
\cos(\psi) & \sin(\psi) & 0 & 0 \cr
-\sin(\psi) & \cos(\psi)  & 0 & 0 \cr
\end{pmatrix},\nonumber\\
&&C_3(\psi)=\begin{pmatrix}
 0 & \cos(\psi) & \sin(\psi) & 0 \cr
 0 & 0 & 0 & 1 \cr
 1 & 0 & 0 & 0 \cr
 0 & -\sin(\psi) & \cos(\psi) & 0\cr
 \end{pmatrix},
C_4(\psi)=\begin{pmatrix}
 0 & 0 & 1 & 0 \cr
 \cos(\psi) & 0 & 0 & \sin(\psi) \cr
 -\sin(\psi) & 0 & 0 & \cos(\psi) \cr
 0 & 1 & 0 & 0\cr
 \end{pmatrix},
\eea
are such that for all realizations $\omega$, $U_\omega(C_j(\psi))$ is pure point.
\end{lem}
{\bf Proof:} Observe that for each $j=1,2,3,4$, the four-dimensional subspaces $\cH_x^j$, labeled by $x\in\cT_4$, satisfy   $\oplus_{x\in\cT_4}\cH_x^j=\cK_4$ and are invariant under $U_\omega(C_j(r))$, where
\bea\nonumber
\cH_x^1&=&\{x\otimes a, xa^{-1}\otimes a^{-1}, xa^{-1}b\otimes b, xa^{-1}\otimes b^{-1}\}\\ \nonumber
\cH_x^2&=&\{x\otimes a, xb^{-1}\otimes b^{-1}, xa^{-1}\otimes a^{-1}, x\otimes b\}\\
\nonumber
\cH_x^3&=&\{x\otimes a, xa^{-1}\otimes a^{-1}, xa^{-1}b^{-1}\otimes b^{-1}, xa^{-1}\otimes b\}\\ \nonumber
\cH_x^4&=&\{x\otimes a, xb\otimes b, xa^{-1}\otimes a^{-1},  x\otimes b^{-1}\}.\hspace{4cm} \ep
\eea
\begin{rem}
The statements (\ref{rpp}), (\ref{pac}) and Lemma \ref{1ppp} remain true if these matrices are decorated by phases, possibly depending on $(\psi, \xi)$.
\end{rem}

\subsection{ Permutation Coin Matrices for $q=4$}
The $24$ permutations of the alphabet  $A_4$ give rise to coin matrices inducing walks with a variety of different  spectral properties. 
A number of them yield fully localizing matrices
\be
\Lambda=\{C_{(abb^{-1}a^{-1})}, C_{(aa^{-1}bb^{-1})}, C_{(aa^{-1}b^{-1}b)}, C_{(ab^{-1}ba^{-1})}, C_{(a a^{-1})(b b^{-1})}\},
\ee
with $C_{(a a^{-1})(bb^{-1})}\in \cR$. 
These matrices are special cases of Lemma \ref{1ppp} and their respective cyclic subspaces labeled by $x\in\cT_4$ are $\cH_x^4, \cH_x^1,  \cH_x^3, \cH_x^2 $ and 
\be
 \cH_x^{12}= {\mbox{span }}\{x\otimes a, xa^{-1}\otimes a^{-1}\}\oplus {\mbox{span }}\{x\otimes b, xb^{-1}\otimes b^{-1}\}.
\ee
There are $9$ permutation coin matrices that are propagating matrices from $\cP$ and give rise to purely absolutely continuous spectrum for any deterministic $\D_\omega$ :
\bea
&&\Pi_1=\{(aba^{-1}b^{-1}), (ab^{-1}a^{-1}b), (ab)(a^{-1}b^{-1}), (ab^{-1})(ba^{-1}),(a)(b)(a^{-1}b^{-1}),  \\ \nonumber
&&\hspace{1.5cm}(a)(b^{-1})(ba^{-1}), (ab^{-1})(b)(a^{-1}), (ab)(a^{-1})(b^{-1}), (a)(b)(a^{-1})(b^{-1}) \}\in \cP.
\eea
These matrices are special cases of $C_j^P(\psi,\xi)$ defined in the previous subsection.
The subset
 \be
\cS=\{C_{(aba^{-1}b^{-1})}, C_{(ab^{-1}a^{-1}b)}\}\subset \Pi_1
\ee
 gives rise to a  spiral-like walk on the tree and are fully delocalizing matrices, see Figure \ref{realspiral}. 
 \begin{figure}[htbp]
   \begin{center}
      \includegraphics[scale=.33]{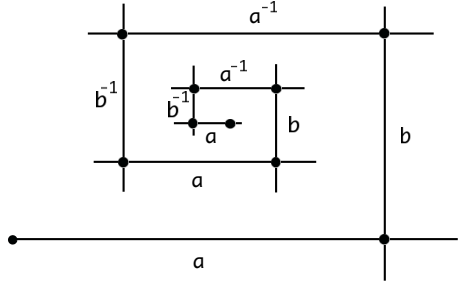}
   \end{center}
   \caption{\footnotesize Spiral-like walk for $(aba^{-1}b^{-1})$.}\label{realspiral}
\end{figure}

All other permutations from $\Pi_1$ give rise to independent shifts on the tree. 
 For example, $U_\omega(C_{(ab)(a^{-1}b^{-1})})$   has
 cyclic subspaces given by
 \bea
\overline{\mbox{span }}\{\dots,xa^{-1}b^{-1}\otimes a, xa^{-1}\otimes b, x\otimes a, xb\otimes b, xba\otimes a,\dots \}
\\ \overline{\mbox{span }}\{\dots,xab\otimes a^{-1}, xa \otimes b^{-1}, x\otimes a^{-1}, xb^{-1}\otimes b^{-1}, xb^{-1}a^{-1}\otimes a^{-1},\dots \}
\eea
labeled with $x\in \cT_4$. 
Another set of permutation matrices that give rise to purely absolutely continuous spectrum but does not belong to $\cP$ is given by
\bea
&&\Pi_2=\{(a)(ba^{-1}b^{-1}), (a)(bb^{-1}a^{-1}), (b)(aa^{-1}b^{-1}),(b)(ab^{-1}a^{-1}),\nonumber \\
&&\hspace{2cm} (a^{-1})(abb^{-1}),  (a^{-1})(ab^{-1}b), (b^{-1})(aba^{-1}), (b^{-1})(aa^{-1}b)\}.
\eea
Let us take a closer look at the operator $U_\omega(C_{(a)(ba^{-1}b^{-1})})$. It leaves the subspace $l^2(\cT_4)\otimes|a\ket$ invariant acting essentially as shifts on the corresponding cyclic subspaces 
\bea
\cH_{x\otimes a}&=&\overline{\mbox{span }}\{\dots,xa^{-1}a^{-1}\otimes a, xa^{-1}\otimes a, x\otimes a, xa\otimes a, xaa\otimes a,\dots \}\\  \nonumber
\cH_{x\otimes b}&=&\overline{\mbox{span }}\{\dots, x\otimes a^{-1},xb^{-1}\otimes b^{-1}, x\otimes b, xa^{-1}\otimes a^{-1}, xa^{-1}b^{-1}\otimes b^{-1}, xa^{-1}\otimes b,\dots \}
\eea
labeled by $x\in\cT_4$, which are easily seen to sum up to $\cK_4$.
The list of permutation matrices  is completed by two coin matrices defining $\cM$ 
\bea
\{C_{(a)(bb^{-1})(a^{-1})}
, C_{(aa^{-1})(b)(b^{-1})}
\}= \cM,
\eea
which are special cases of $C^R(\psi,\xi)$ defined in (\ref{defred}).
A closer look at $U_\omega(C_{(a)(bb^{-1})(a^{-1})})$ shows that it leaves the subspaces $l^2(\cT_4)\otimes| a\ket$ and $l^2(\cT_4)\otimes |a^{-1}\ket$ invariant, where the dynamics is essentially driven by shift $S_a$ and  $S^{-1}_a$ acting on the cyclic subspaces $\overline{\mbox{span }}\{\cdots,xa^{-1}\otimes a,x\otimes a,xa\otimes a,\cdots\}$  and $\overline{\mbox{span }}\{\cdots,xa\otimes a^{-1}, x\otimes a^{-1},xa^{-1}\otimes a^{-1},\cdots\}$ labeled by $x\in\cT_4$. On the other hand, for all $x\in\cT_4$, the two dimensional subspace 
$\overline{\mbox{span }}\{x\otimes b, xb^{-1}\otimes b^{-1}\}$ is invariant under $U_\omega(C_{(a)(bb^{-1})(a^{-1})})$. Therefore the spectrum contains both absolutely continuous and pure point parts. The case of $ C_{(aa^{-1})(b)(b^{-1})}$ is similar.

\begin{rem} All results of this section hold true if the permutation matrices $C_\pi\in U(4)$ are replaced by decorated permutation matrices $ C^\Phi_\pi$.
\end{rem}

\subsection{Spectral Transition for $q=4$}

\begin{figure}[!h]
   \begin{center}
      \includegraphics[scale=.38]{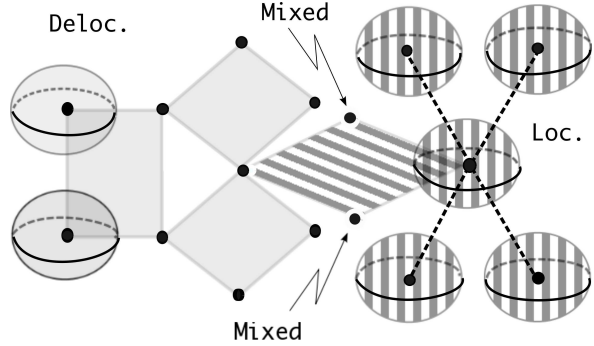}
   \end{center}
   \caption{\footnotesize Partial spectral diagram for $q=4$.}\label{phase4}
\end{figure}

We showed the existence of a continuous path of coin matrices which links localizing matrices from a small neighborhood of $\Lambda$ to delocalizing matrices from a small neighborhood of $\cS$. 
All elements of $\Lambda$ are linked by paths described by the one parameter families $C_j(\psi)$, with suitable decorating phases, giving rise to pure point spectrum for all $\omega$. Then $C_{(aa^{-1})(bb^{-1})}\in\Lambda$ is linked to $C_{(a)(b)(a^{-1})(b^{-1})}\in \Pi_1\subset \cP$ by the two parameter family $C^R(\psi,\xi)$ with suitable decorating phases, which gives rise to pure point spectrum, for almost all $\omega$. Eventually, $C_{(a)(b)(a^{-1})(b^{-1})}$ is linked to all other elements of $\Pi_1$, in particular to the elements of $\cS$, by the  the two parameter families $C_j^P(\psi,\xi)$ with suitable decorating phases which yield absolutely continuous spectrum for all $\omega$. This is illustrated in Figure \ref{phase4}, where the elements of $\Pi_2$ do not appear since they play no role in this transition.

\medskip
\appendix
\section{ Proof of Theorem \ref{fme}:}
While the result holds for any $q$ and any permutation matrix in $\Lambda$, we provide a  proof for $q$ odd and for $\pi_o=(1 2 \dots q)$ only. The case $q$ even is somehow simpler whereas the modifications required by different choices of fully localizing permutations are dealt with along the lines of \cite{J3}.  
In the following, the symbol $c$ denotes unessential constants, that may vary from line to line. 
The first step towards (\ref{am}) is an 
estimate on  fractional moments of the finite volume Green function, 
\begin{prop}\label{finvolam}
For all 
$0 < s < 1$,  all $p'>1/(1-s)$ fixed, there exist $C(s)$ and $c_0(s)<\infty$ so that for all $\alpha> 0$, all $C\in U(q)$ such that $\|C-C^\Phi_{\pi_0}\|\leq c_0(s)e^{-L\alpha(1/s+2/p')}(q-1)^{-2L}$, the estimate
\be\label{fvam} 
\E(|G^{\Lambda_L(x_e)}_{a_j,a_k,\omega}(x,y;C,{z})|^s)\leq C(s)e^{-\alpha L},
\ee
holds for all $L\geq 3$, all $z\not\in \U$, all $x\otimes a_j,y\otimes a_k\in \cH^{\Lambda_L(x_e)}$ with $d(x,y) >2$. \\ The estimate also holds for $\alpha=0$, without restriction on $C$ or $d(x,y)$.
\end{prop}
{\bf Proof:}
We first note that Theorem 3.1 of \cite{HJS2} gives the required estimate for $\alpha=0$,
\be\label{exgreen}
\E(|G_{a_j,a_k,\omega}(x,y;C,{z})|^s)\leq C(s).
\ee
The desired exponential decay follows from the second resolvent identity  and perturbation theory as in \cite{J3, ABJ2, HJS2}, taking care of the dependence in $q$ of the estimates.
 \ep \\
The second step makes the link between estimates on finite and infinite volume Green functions.
We drop the symbols $C$, $\omega$, $x_e$ and $z\not\in \U$ in the notation. We set $T^L$ by 
\be\label{decomp}
U=U^L+T^L=U^{\Lambda_L} \oplus U^{\Lambda_L^c} + T^{L},
\ee
see Lemma \ref{last}, and we keep track of the dependence in $t=\|T^L\|$, where $t\leq c \|C-C^\Phi_{\pi_0}\|$, uniformly in $L$ and $\omega$.
We denote by $G^L_{a_j,a_k}(x,y)$ the Green function corresponding to 
$
 G^{L}= (U^L-z)^{-1} = (U^{\Lambda_L} \oplus U^{\Lambda_L^c}-z)^{-1} = (U^{\Lambda_L}-z)^{-1} \oplus (U^{\Lambda_L^c}-z)^{-1}
 $.
 \begin{prop}\label{itsteprop} 
 For every $s\in (0,1/3)$ there exists a constant $c_1(s) <\infty$ depending on $s$ (and $q$), such that
\bea \label{itstep}
&&\E\left(\vert G_{a_j,a_k}(x,y)\vert^{s}\right)\leq c_1(s) t^{2s}  (1+c_1(s) t^s(q-1)^{L} )\\ \nonumber
&&\hspace{1.cm}\times \sum_{u \in \cH \atop  |d(u_1,x_e)-L|\leq 2} \E\left(\vert G^{L}_{a_j,u_2}(x,u_1)\vert^{s}\right)\sum_{x' \in\cH\atop  |d(x'_1,x_e)-(L+3)|\leq 2}\E\left(\vert G_{x'_2,a_k}(x'_1,y)\vert^{s}\right)
\eea
uniformly in $z\not\in\U$ with $1/2< |z| < 2$, $L\in \N$ and $x, y\in \cT_q$ with $d(x,x_e)\leq L$ and $d(y,x_e)> L+5$, with the notation $u=u_1\otimes u_2\in \cH$, $u_1\in \cT_q$, $u_2\in A_q$
\end{prop}
{\bf Proof:}
We use a resampling argument to decouple the expectations, then the general estimate (\ref{exgreen}) to get rid of the full resolvent term. 
This step requires dealing with the metric peculiarities of the tree. In particular, the estimate, for $k, j $ fixed,
$
\#\{w \in \cH \  |d(w_1,x_e) -(L+k)|\leq L+j \}\leq c(k,j)(q-1)^L
$, 
eventually yields (\ref{itstep}), similarly to what is done to get Proposition 13.2 in \cite{HJS2}. \ep

Finally, one uses an iterative argument to eventually reach  (\ref{fme}), taking care of the dependence in $(s,\alpha)$ of the different parameters, considering $q$ fixed.
We first note that  thanks to (\ref{fvam}), for $L\geq 3$ and for some $C_q(s)$,
\be\label{65}
\sum_{u \in \cH \atop  |d(u_1,x_e)-L|\leq 2} \E\left(\vert G^{L}_{a_j,u_2}(x,u_1)\vert^{s}\right)\leq C_q(s)(q-1)^Le^{-\alpha L},
\ee
if $d(x,x_e)\leq L$.  Our hypothesis on the perturbation $\|C-C^\Phi_{\pi_0}\|$ with $t\leq c\|C-C^\Phi_{\pi_0}\|$ implies for any $p'>1/(1-s)$ and some $c_q(s)<\infty$,
\be
t^s (q-1)^L\leq c_q(s) e^{-L\beta(\alpha)}, \mbox{ with }\ \beta(\alpha)= \alpha(1+2s/p') -\ln(q-1)(1-2s). 
\ee
Hence, due to (\ref{fvam}) and (\ref{65}), given $0<s<1/3$ and $p'>1/(1-s)$, there exists $\alpha_0(s)>0$ (depending on $q$ and $p'$) such that, for all $x, y\in \cT_q$ with $d(x,x_e)\leq L$, $d(y,x_e)>L+5$, and $\alpha\geq \alpha_0(s)>0$ we have $\beta(\alpha)>0$ and for some ($q$ and $p'$ dependent) $c(s)<\infty$ 
\bea \label{startiter}
&&\E\left(\vert G_{a_j,a_k}(x,y)\vert^{s}\right)\leq c(s) e^{-L(\alpha(3+ 4s/p')-(1-4s)\ln(q-1))}  (1+c(s) e^{-\beta(\alpha) L} )\\ \nonumber
&&\hspace{2.cm}\times \sup_{x' \in\cH\atop  |d(x'_1,x_e)-(L+3)|\leq 2}\E\left(\vert G_{x'_2,a_k}(x'_1,y)\vert^{s}\right)\\ \nonumber
&&\hspace{2.cm}
\leq c_0(s) e^{-L\delta(\alpha)}\sup_{x' \in\cH\atop  |d(x'_1,x_e)-(L+3)|\leq 2}\E\left(\vert G_{x'_2,a_k}(x'_1,y)\vert^{s}\right),  \ \mbox{with}
\eea
\be
\delta(\alpha)=\alpha(3+ 4s/p')-(1-4s)\ln(q-1)\geq \alpha_0(s)(3+ 4s/p')-(1-4s)\ln(q-1)>0.
\ee
Now let
$
b(s,\alpha)=c_0(s) e^{-L_0(s)\delta(\alpha)}
$, 
and fix $L_0(s)=L_0(s,\alpha_0(s))$ odd and large enough so that $b(s,\alpha_0)<1$ .
Thus for any $\alpha\geq \alpha_0(s)$, $b(s,\alpha)\leq b(s,\alpha_0(s))<1$. This determines the size of the perturbation via
\be
\|C-C^\Phi_{\pi_0}\|\leq \Delta(s,\alpha):= c_0(s)e^{-L_0(s)\alpha(1/s+2/p')}(q-1)^{-2L_0(s)}\leq \Delta(s,\alpha_0(s)).
\ee 
By ergodicity, see (\ref{ergodic}), 
$
\max_{a_j,a_k\in A_q}\E\left(\vert G_{a_j,a_k}(x,y)\vert^{s}\right)=\max_{a_j,a_k\in A_q}\E\left(\vert G_{a_j,a_k}(x',y')\vert^{s}\right)
$
for all $x'=zx, y'=zy \in \cT_q$ with $|z|$ even,  where $d(x',y')=d(x,y)$. Thus, in the right hand side of (\ref{startiter}), we can shift the arguments of the Green function so that $x_1'$ is equal or close to the center of the ball $\Lambda(x_e)$ and provided $d(x_1',y)\geq L+5$ one can iterate (\ref{startiter}). Doing this along a sequence of points forming a path of length of order $d(x,y)=n_y L_0$, we get  
that 
\be
\E\left(\vert G_{a_j,a_k}(x,y)\vert^{s}\right)\leq c b^{n_y}(s,\alpha)\leq c e^{-\gamma(\alpha) d(x,y) },
\ee
where, for $\alpha$ large enough,
\be
\gamma(\alpha) = \delta(\alpha) - \ln(c_0)/L_0(s)=\alpha(3+ 4s/p')-(1-4s)\ln(q-1)- \ln(c_0)/L_0(s) >0.
\ee  Since $\gamma(\alpha) $ is invertible and can be made arbitrarily large by increasing $\alpha$, we get the result by defining $\eps(s,\gamma)=\Delta(s,\alpha^{-1}(\gamma))$.   \ep

 {

}

\begin{thebibliography}{99}
%
\bibitem[AAKV]{AAKV} D.\ Aharonov, A.\ Ambainis, J. Kempe, U. Vazirani, Quantum Walks on Graphs, {\it STOC 2001} Proc. of the 33rd ACM symposium on Theory of computing, 50-59 (2001).
%
\bibitem[ASWe]{ASWe} A.\ Ahlbrecht, V.B.\ Scholz, \ A.H.\ Werner, Disordered quantum walks in one lattice dimension, {\it J. Math. Phys.}, {\bf  52}, 102201 (2011).
%
\bibitem[AVWW]{AVWW} A.\ Ahlbrecht, H.\ Vogts, A.H.\ Werner, and R.F. \ Werner, Asymptotic evolution of quantum walks with random coin,  {\it J. Math. Phys.}, {\bf 52}, 042201 (2011).
%
\bibitem[A]{A} M. Aizenman: Localization at weak disorder: Some elementary bounds. {\it Rev. Math. Phys.} {\bf }6, 1163-1182 (1994).
%
\bibitem[AM]{AM} M.\ Aizenman, S.\  Molchanov, Localization at  large disorder
and at extreme energies: an elementary derivation, {\it Commun. Math. Phys.}
{\bf 157}, 245-278, (1993).
%
\bibitem[ASWa]{ASW} M. Aizenman, B. Sims, S. Warzel, Stability of the absolutely continuous spectrum of random Schr�dinger operators on tree graphs, {\it Probab. Theory Relat. Fields} {\bf 136}, 363-394 (2006)
%
\bibitem[AW1]{AW1} M. Aizenman, S. Warzel, Absence of mobility edge for the Anderson random potential on tree graphs at weak disorder, {\em EPL} {\bf 96} 37004, (2011).
%
\bibitem[AW2]{AW2} M. Aizenman, S. Warzel, Resonant delocalization for random Schr�dinger operators on tree graphs, arxiv 1104.0969, (2011).
%
\bibitem[ABJ]{ABJ2}J.\ Asch , O.\ Bourget and A.\ Joye, Dynamical Localization of the Chalker-Coddington Model far from Transition, {\it J. Stat. Phys.}, {\bf 147}, 194-205 (2012).
%
\bibitem[APSS]{APSS}S. \ Attal, F. \ Petruccione, C. \ Sabot, I. \ Sinayski. Open Quantum Random Walks,  {\it J. Stat. Phys.}, {\bf 147}, 832-852 (2012).
%
\bibitem[BHJ]{BHJ} O.\ Bourget J. S. \ Howland,  A.\ Joye, Spectral Analysis of Unitary Band Matrices,
{\em Commun. Math. Phys.}, {\bf 234}, (2003), p. 191-227
%
\bibitem[CC]{CC} Chalker, J.T., Coddington, P.D.: Percolation, quantum tunneling and the integer Hall effect, {\it J. Phys. C} {\bf 21}, 2665-2679, (1988).
%
\bibitem[CHKS]{CHKS} K. Chisaki, M. Hamada,	N. Konno,	E. Segawa, Limit theorems for discrete-time quantum walks on trees, {\em Interdisciplinary Information Sciences}, {\bf 15},423--429, (2009).
%
\bibitem[D et al]{Detal} Dimcovic, Z., Rockwell, D., Milligan, I., Burton, R. M., Nguyen, T.,  Kovchegov, Y., Framework for discrete-time quantum walks and a symmetric walk on a binary tree, {\em Phys. Rev. A}, {\bf 84}, 032311, (2011).
%
\bibitem[Gu]{Gu} S. Gudder, Quantum Markov Chains, {\em J. Math. Phys.}, {\bf 49}, 072105, (2008).
%
\bibitem[HJ]{HJ} E.\ Hamza, A.\ Joye, Correlated Markov Quantum Walks, {\it Ann.\ H.\ Poincar\'e}, {\bf 13}, 1767-1805, (2012).
%
\bibitem[HJS1]{HJS1} E.\ Hamza, A.\ Joye and G.\ Stolz,  Localization for Random Unitary Operators, 
{\em Letters in Math. Phys.}, {\bf 75}, (2006), p. 255-272 .
%
\bibitem[HJS2]{HJS2} E.\ Hamza, A.\ Joye and G.\ Stolz, Dynamical Localization for Unitary Anderson Models, 
{\it Math. Phys., Anal. Geom.}, {\bf 12}, 381-444 (2009).
%
\bibitem[J1]{J1} A.\ Joye, Fractional moment estimates for random unitary operators, {\it Lett. Math. Phys.} {\bf 72}, no. 1, 51--64 (2005).
%
\bibitem[J2]{J2} A.\ Joye, Random Time-Dependent Quantum Walks, {\it Commun. Math. Phys.}, {\bf 307}, 65-100 (2011).
%
\bibitem[J3]{J3} A.\ Joye, Dynamical Localization for $d$-Dimensional Random Quantum Walks, {\em Quantum Inf. Process.}, ,  Special Issue: Quantum Walks, {\bf 11}, 1251-1269,  (2012).
%
\bibitem[J4]{J4} A.\ Joye,  Dynamical Localization of Random Quantum Walks on the Lattice, {\em Proc. of the ICMP}, August 6-11th, Aalborg (2012). To appear.
%
\bibitem[JM]{JM} A.\ Joye, M. \ Merkli, Dynamical Localization of Quantum Walks in Random Environments, {\it J. Stat. Phys.}, {\bf 140}, 1025-1053, (2010).
%
\bibitem[Ke]{Ke} J.\ Kempe, Quantum random walks - an introductory overview, {\it Contemp.  Phys.}, {\bf 44}, 307-327, (2003).
%
\bibitem[K et al]{Ketal} M.\ Karski, L.\ F\"orster, J.M. \ Chioi, A. \ Streffen, W.\ Alt, D.\ Meschede, A. \ Widera,  Quantum Walk in Position Space with Single Optically Trapped Atoms, {\it Science}, {\bf 325}, 174-177, (2009).
%
\bibitem[KLMW]{KLMW} J. P.\ Keating, N.\ Linden, J. C. F.\ Matthews, and A.\ Winter,
Localization and its consequences for quantum walk algorithms and quantum communication, {\it Phys. Rev. A} {\bf 76}, 012315 (2007).
%
\bibitem[Kl]{Kl} A. Klein, Extended states in the Anderson model on the Bethe lattice, {\it Adv. Math.}, {\bf 133}, 163�184, (1998)
%
\bibitem[Ko]{Ko} N.\ Konno, Quantum Walks, in "Quantum Potential Theory", Franz, Sch\"urmann Edts, {\it Lecture Notes in Mathematics}, {\bf 1954}, 309-452,  (2009).
%
\bibitem[MNRS]{MNRS} F. Magniez, A. Nayak, J. Roland, and M. Santha, Search via quantum walk. {\em SIAM Journal on Computing}, {\bf 40}, 142-164, (2011).
%
\bibitem[RS]{RS} M. Reed, B. Simon, Methods of Modern Mathematical Physics, Vol. 3, Academic Press, 1979.
%
\bibitem[S]{S} M. Santha, Quantum walk based search algorithms, 5th TAMC, {\it LNCS} {\bf 4978}, 31-46, 2008.
%
\bibitem[S et al]{sciarrino} S. Spagnolo,  C. Vitelli, L. Aparo, P. Mataloni, F. Sciarrino,  A. Crespi, R. Ramponi, R. Osellame, Three-photon bosonic coalescence in an integrated tritter, arxiv 1210.6935, 2012
%
\bibitem[V-A]{V-A} Venegas-Andraca, Salvador Elias, Quantum walks: a comprehensive review, {\em Quantum Inf. Process.},  {\bf 11}, 1015-1106, (2012).
%
\bibitem[W]{W} W. Woess, Generating function techniques for random walks on graphs, {\em Contemporary Mathematics}, {\bf 338}, 391�423, (2003).
%
\bibitem[Z et al]{Zetal} F.\ Z\"ahringer, G.\ Kirchmair, R.\ Gerritsma, E.\ Solano, R.\ Blatt, C. F.\ Roos, Realization of a quantum walk with one and two trapped ions, {\it Phys. Rev. Lett. } 104, 100503 (2010).

\end{thebibliography}
\end{document}